\documentclass[10pt]{article}
\usepackage{euscript,amsmath,amssymb,amsfonts,graphicx,bm}
\usepackage[justification=justified,format=plain]{caption}
  \usepackage{paralist}
  \usepackage{epstopdf}
  \usepackage{graphics} 
    \usepackage{amsmath} 
  \usepackage{enumitem}
 \usepackage[colorlinks=true]{hyperref}
 \hypersetup{urlcolor=black, citecolor=black}
 \usepackage{xcolor}
\usepackage{natbib}
\setcitestyle{authoryear,open={(},close={)}}
\allowdisplaybreaks
 
\usepackage{changepage}

\usepackage[utf8x]{inputenc}

\usepackage{textcomp,marvosym}
 
\newcommand{\e}{{\rm e}}

\newcommand{\bi}[1]{\textbf{\em #1}}

\usepackage[aboveskip=1pt,labelfont=bf,labelsep=period,justification=raggedright,singlelinecheck=off]{caption}

\date{}

\usepackage{lastpage,fancyhdr,graphicx}
\usepackage{epstopdf}

\setlist[itemize]{leftmargin=*}

\begin{document}
\vspace*{0.2in}

\begin{flushleft}

{\Large \bf Hive geometry shapes the recruitment rate of honeybee colonies}
\newline
\\
\bigskip
Subekshya Bidari\textsuperscript{1} and Zachary P Kilpatrick\textsuperscript{1,*} 
\\
\bigskip
\textbf{1} Department of Applied Mathematics, University of Colorado, Boulder CO, USA
\\
\textbf{*} zpkilpat@colorado.edu
\end{flushleft}

\section*{Abstract}
Honey bees make decisions regarding foraging and nest-site selection in groups ranging from hundreds to thousands of individuals. To effectively make these decisions bees need to communicate within a spatially distributed group. However, the spatiotemporal dynamics of honey bee communication have been mostly overlooked in models of collective decisions, focusing primarily on mean field models of opinion dynamics. We analyze how the spatial properties of the nest or hive, and the movement of individuals with different {\em belief states} (uncommitted or committed)  therein affect the rate of information transmission using spatially-extended models of collective decision-making within a hive. Honeybees waggle-dance to recruit conspecifics with an intensity that is a threshold nonlinear function of the waggler concentration. Our models range from treating the hive as a chain of discrete patches to a continuous line (long narrow hive). The combination of population-thresholded recruitment and compartmentalized populations generates tradeoffs between rapid information propagation with strong population dispersal and recruitment failures resulting from excessive population diffusion and also creates an effective colony-level signal-detection mechanism whereby recruitment to low quality objectives is blocked. \\[1ex]
\noindent
{\bf Keywords:} collective decision-making, foraging, optimality, social insects, dynamic environments


\section{Introduction} 

Honeybees forage in groups, as many animals do~\citep{clark86,krause02}. Groups of hundreds to thousands of foragers work together in a colony to explore large patches of flowers, allowing foragers to focus on the richest flower patches~\citep{seeley86}. Collective foraging in social insects emerges through group members' interactions in a decentralized decision-making process~\citep{camazine03,detrain06}. In a honey bee colony, collective choice recruitment arises through the waggle dance, a communication behavior by which colony members share information about the locations of profitable food sources~\citep{frisch1967,seeley88}. Worker bees perform a figure-eight dance after returning to the hive from foraging to indicate the direction and distance to high-quality foraging patches~\citep{gruter09,seeley10}. Communication through waggle dancing enables a honey bee colony to explore large patches of flowers and collectively narrow their choice to the richest flower patches to forage~\citep{seeley86}.

Spatial properties of nest or hive impact the collective movement and decisions of social insects inside~\citep{gruter06,burd10}. It is well known that the geometry and topology of networks strongly impact the rate of information transmission along them~\citep{karsai11,mateo19}, but there has been little quantitative work addressing this aspect of communication in honeybee hives. We study here how the spatial structure of a honeybee hive affects colony-wide recruitment as bees share information via waggle dancing. In an enclosed hive, the movement of bees and presence of wagglers determines whether an uncommitted bee in a nest is recruited to forage. The speed at which the wagglers and the uncommitted bees move is determined by the hive's geometry, parameterized in our model by the number of hive compartments and diffusion rate between them. Indeed, previous studies have shown that in ant colonies, the probability an ant encounters another ant depends on the nest architecture~\citep{razin13,davidson16,pless15}. 

Given undecided and decided agents are in proximity, classic models of collective decisions often use mass action principles to define opinion dynamics within the group~\citep{franks02,seeley12}. The rate of recruitment scales linearly with the density of group mates expressing opinions. However, decision commitment in social insects (e.g., bees and ants) may be better fit by a nonlinear function of choice evidence~\citep{bonabeau96,mailleux06}. When decision commitment depends on a threshold function of choice evidence, deleterious recruitment to overvalued choices is curbed~\citep{pagliara18}. Moreover, key features of individualized cognition (like threshold-dependent neural activation) are also present in the collective cognition of honeybee groups~\citep{passino08}. As such, we extend previous models of collective decisions in honeybees, considering waggle-based recruitment that is a threshold nonlinear function of the concentration of wagglers. Combining thresholded recruitment and a spatially-extended hive geometry makes for a more robust recruitment process, limiting loss from erroneous recruitment while ensuring speedy recruitment when warranted.

Our model of recruitment assumes a bee waggle dances to recruit nest mates to her opinion. Rather than considering choices between multiple foraging or house-hunting alternatives~\citep{seeley12,reina17}, the hive has a decision between departing to an external choice (forage) or remaining in the hive. The movement of honey bees in the hive shaped by the architecture of the hive determines the rate of information transfer between bees. When bees diffuse quickly through the hive, they can facilitate rapid information transfer in the colony. However, colonies might not always benefit from rapid information transfer, especially when the information is erroneous. In our previous work, we showed that in a volatile environment with some low quality resources, bees are often better off using private information instead of the social information available through the beliefs of neighbors~\citep{price19, bidari19}. Indeed, positive feedback from waggle dancing not only generates consensus but can also amplify errors~\citep{sasaki18}. We demonstrate this by analyzing our models in several tractable scenarios which exploit a separation of timescales in the limit of slow/fast movement dynamics, and due to the piecewise linearity of simple threshold functions.

The combination of thresholded recruitment and compartmentalized population distributions within the hive allows the colony recruitment process to behave as a switch. When evidence for a high quality option outside the hive is plentiful, there will be many wagglers which will exceed the recruitment threshold of uncommitted bees. However, if there is an insufficient fraction of the hive waggling or wagglers diffuse quickly throughout the hive, no compartments of the hive will have sufficient wagglers to initiate recruitment. It is therefore imperative that the colony balance the advantages of diffusing throughout the hive to quickly spread a recruitment message with the limitations in diffusing too fast, which can result in failed collective recruitment. Our model analysis therefore leverages tools from nonlinear dynamics to determine the parametric optima that appropriately balance these tradeoffs, in the process shedding light on the functional advantages of a spatially-distributed hive subject to thresholded recruitment.

\section{Collective decision model with threshold recruitment} 

\begin{figure}[h]
\begin{center}    \includegraphics[width=15cm]{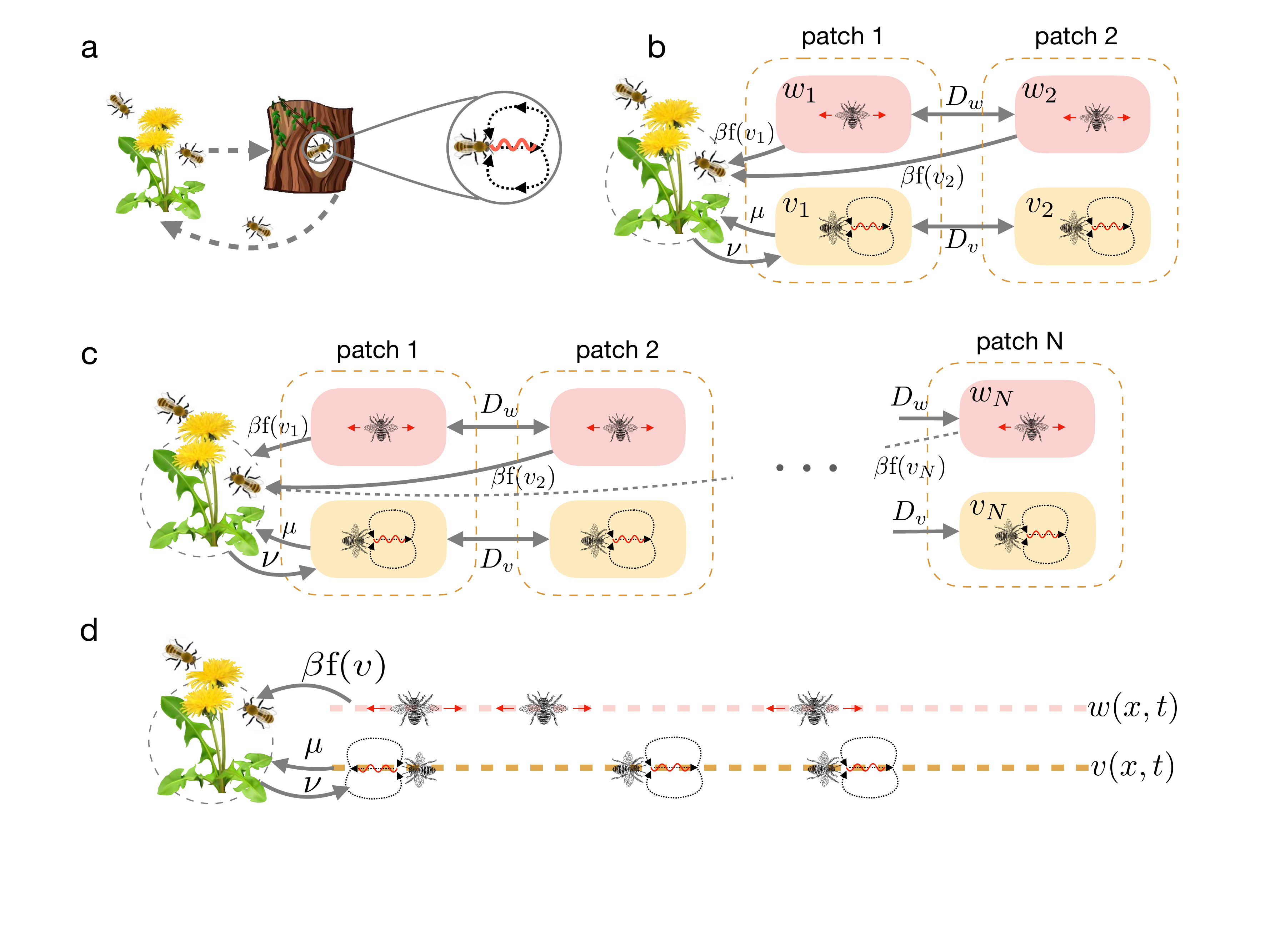}	\end{center} 	
\caption{{\bf Schematics of how hive elongation shapes bee movement and recruitment.}  {\bf a.} Waggling bee recruits neighbors to a foraging site; {\bf b.} Two patch model in which uncommitted bees $w_j$ can only be recruited by wagglers in the same patch $v_j$ ($j = 1,2$); {\bf c.} $N$ patch hive requires recruiters originating in patch 1 ($v_1$) to diffuse across the hive to recruit uncommitted bees in distal patches; {\bf d.} Continuum model: Bees move continuously via diffusion along a one-dimensional line segment.}
\label{fig1}
\end{figure}

We consider a model of recruitment in a honey bee colony that applies to foraging and nest site colonization, extending prior work which ignored spatial aspects of the communication process~\citep{franks02,seeley12}. The analysis of this paper specifically focuses on bees solving a foraging problem. A hive consists of $N$ discrete compartments with a total number of bees. The population is split into fractions that are foraging $u$ (\textit{actively collecting nectar at flower patches}), waggling  $v$ (\textit{in the hive waggling to recruit uncommitted bees}), or uncommitted $w$ (\textit{uncommitted and in the hive}). The mean field model we analyze evolves according to the following system of equations (See also schematics in Fig.~\ref{fig1}):
\begin{subequations} 	\begin{align}
\frac{\partial w_j}{\partial t}  =&  - \beta H[ v_j-\theta] w_i +  \begin{cases}  D_w(w_{j \pm 1}-w_j), \quad & j=1,N \\ D_w(w_{j+1}-2 w_j + w_{j-1}), & \text{else} \end{cases} \\
\frac{\partial v_j}{\partial t}  =& -\mu v_j + \begin{cases}  \nu u  + D_v(v_{j + 1}-v_j), \quad & j=1 \\ D_v(v_{j - 1}-v_i),   & j = N \\  D_v(v_{j+1}-2 v_j + v_{j-1}), & \text{else} \end{cases} \\
\frac{\partial u}{\partial t}  =& - \nu u + \beta \sum_{j} H[ v_j-\theta] w_j + \mu \sum_{j} v_j,
\end{align}\label{discreteEqn}
\end{subequations}	
where movement within the hive is characterized by the diffusion coefficients $D_w$ and $D_v$ of uncommitted bees and wagglers. Conservation ensures $u(t) + \sum_{j=1}^N v_j(t) + w_j(t) = 1$ and $u'(t) + \sum_{j=1}^N v_j'(t) + w_j'(t) = 0$. We assume spontaneous commitment typical of such collective decision models~\citep{franks02,seeley12} is weaker than recruitment and does not influence short-term recruitment dynamics considered here. Likewise, we omit spontaneous abandonment term, as bees are not competing against different foraging sites and the inclusion of abandonment term would simply avert commitment of some of the bees in the hive. Uncommitted bees in each compartment $j$ are recruited to forage according to a rate $\beta$ and threshold function $H(v_j - \theta)$ of the fraction of wagglers $v_j$ in that compartment. As we show, thresholding can prevent erroneous recruitment to low yielding or perilous foraging sites. Foragers switch to becoming wagglers at a rate $\nu$ and switch back to foraging at a rate $\mu$. Note in Eq.~(\ref{discreteEqn}), we assume waggling bees enter the hive at the first compartment ($E = 1$: on the left end of the hive), but we will also analyze cases in which the entrance is in the middle ($1< E< N$). 

Bees are assumed to be homogeneously mixed within each compartment, and compartments are small enough such that all bees within one compartment can perceive a waggler within the compartment but not wagglers in other compartments. Our analysis begins by focusing on the simplest case of two hive compartments (Fig.~\ref{fig1}b) and subsequently increases the number of hive patches (Fig.~\ref{fig1}c) before moving to the continuum limit (Fig.~\ref{fig1}d). 

\section{Compartmental hives threshold recruitment events}

In a spatially-extended model of a hive, the movement of uncommitted and committed individuals affects the rate of information transmission. Slow information transfer leads to lags in foraging recruitment. However, if individuals disperse throughout the hive too quickly, clusters of recruiters may not be large enough to initiate population-wide recruiting. Using numerical simulations, quasi-steady state analysis, and bounding arguments, we show there exists a critical level of recruiter diffusion that allows the colony to balance the tradeoffs between rapid information propagation resulting from population movement with recruitment failures resulting from excessive population diffusion, which spreads the recruiting signal too thin. Our derivation of the critical diffusion level is aided by the tractability of the piecewise linear model of recruitment communication.

Since recruitment initiation depends on a threshold function of recruiter density, diffusion of recruiters across the hive can eventually lead to all local densities of recruiters falling below threshold, even when some regions begin superthreshold. Note this is not due to a loss of recruiters in the overall population, but to the fraction per patch thinning as the population spreads. This creates an effective colony-level signal-detection mechanism that can block recruitment to low quality foraging sites when an insufficient number of recruiters attempts to initiate foraging in the rest of the colony. We will later show how, using signal detection theory, we can tune the recruitment threshold to maximize the foraging yield of a colony across possible environments (favorable/unfavorable to foraging).

In our analysis of the efficacy of population-wide recruiting, we use two measures to illustrate how hive geometry shapes recruitment: {\em i. rate of information transfer}, and {\em ii. foraging yield}. We quantify the rate of information transfer using the time taken to recruit $90 \%$ of the hive population, $T_{90}$ where
\begin{align*}
u(T_{90}) + \sum_{i=1}^N v_i(T_{90}) = 0.90.	
\end{align*}
Here, $T_{90}$ represents the time when nearly all of the hive is recruited. Our results do not change qualitatively when considering the time to commit $80 \%$ or $95 \%$ of the population instead.

Foraging yield refers to the total food (e.g., nectar) collected in a foraging cycle. We assume the waggle dancing bees forage between waggle runs in the hive while {\em active foragers} mostly forage but do not waggle dance.  We calculate foraging yield using
\begin{align*}	
J(\beta,D_v,D_w,\nu,\mu) = \int_{0}^{T_f} r(t) \ dt	
\end{align*}
where $r(t) = u(t) + \gamma \sum_{j=1}^N v_j(t)$ is comprised of food reward provided by active foragers $u(t)$ and wagglers $v_j$, which is a fraction $\gamma \in (0,1)$ of that provided by active forager. Colonies can maximize their foraging yield by increasing their recruitment rate $\beta$. However, a high recruitment rate could be detrimental to a colony in a rapidly changing environment~\citep{pagliara18,bidari19}. Alternatively, colonies may tune their diffusion rates ($D_v, D_w$) for instance by modifying the interior shape of their hive, which controls the rate of rapid information propagation about the hive, balancing higher foraging yields with restrained recruitment in low quality environments.

\subsection{Recruitment dynamics is shaped by the movement of waggling bees}

We begin by considering a hive with two compartments. As we ignore spontaneous commitment and abandonment common in decision models for honey bees, uncommitted bees are initially recruited to forage by a fraction of waggle dancing bees in the first hive patch with density $v_1(0)$, which could be scout bees who have identified a potential foraging site~\citep{biesmeijer05}. Aside from the initial fraction of foragers $u(0)$, we assume the bees remaining in the hive are uncommitted and spread evenly ($w_1(0) =w_2(0)= \frac{1-v_1(0) - u(0)}{2} \equiv \bar{w}(0)$).
Note, we take $v_2(0) = 0$, indicating a group of recruiters arrives through an entrance in the first patch.

If $\phi \equiv v_1(0) + u(0)< \theta$ initially, it is straightforward to show that no recruitment occurs. To see this, note that clearly $H(v_j(t) - \theta) = 0$ near $t=0$, so  Eq.~(\ref{discreteEqn}) reduces to $w_1(t) = \bar{w} (0)  = w_2(t)$ and
\begin{align*}
\dot{v}_1 = - \mu v_1 + D_v(v_2 - v_1) + \nu u; \ \ \dot{v}_2 = - \mu v_2 + D_v(v_1 - v_2); \ \ \dot{u} = - \nu u + \mu (v_1 + v_2).
\end{align*}
Clearly, $v_1 + v_2 + u = \phi $, since $\dot{v}_1 + \dot{v}_2 + \dot{u} = 0$. Moreover, the rate of change of each density $(v_1, v_2, u)$ is non-negative when $v_1= 0$ or $v_2=0$ or $u = 0$ and the other densities are non-negative, that is
\begin{align*}
\dot{v}_1 &= D_v v_2 + \nu u \geq 0, \ {\rm if} \ v_1 =0 \ \& \ v_2, u \geq 0; \\
\dot{v}_2 &= D_v v_1 \geq 0, \ {\rm if} \ v_2 = 0 \ \& \ v_1 \geq 0; \\
\dot{u} &= \mu (v_1 + v_2) \geq 0, \ {\rm if} \ u = 0 \ \& v_1, v_2 \geq 0,
\end{align*}
implying $v_1, v_2, u \geq 0$ for all time if $v_1(0), v_2(0), u(0) \geq 0$. Together with the conservation condition, this implies $v_1, v_2, u \leq \phi < \theta$, so recruitment will never begin.
Thus, the diffusive coupling between patches equilibrates the distribution of wagglers in the hive over time, so as long as $v_1(0)>v_2(0)$, then $v_1(t)$ will decrease and neither patch density can exceed $v_1(0)$, implying $u(0) + v_1(0)<\theta$ leads to no recruitment.

To examine the dynamics of recruitment, we must assume $u(0) + v_1 (0) > \theta$, in which case there are three possible phases of colony behavior determined by initial conditions and model parameters. We defer the analysis of the full model in Eq.~(\ref{discreteEqn}) to the later in this section and start by considering limiting case of $\nu \to \infty$ and $\mu \to 0$ so all recruits become wagglers. We also consider the limits of committed ($D_v \to 0$ and $D_v \to \infty$) and uncommitted ($D_w \to 0$ and $D_w \to \infty$) bee movements as well as slow and fast recruitment ($\beta \to 0$ and $\beta \to \infty$) to demonstrate how changes in recruitment rate and diffusion impact qualitative recruitment behavior within the hive.

\subsubsection{Limiting cases of no foragers}

By assuming $\nu \to \infty$ and $\mu \to 0$, any committed bee instantaneously becomes a recruiter for the site. To see this, note that in this limit, $u \to 0$, so that
\begin{align*}
v_1'(t) + u'(t) = v_1'(t) = \beta \left[ H[v_1 - \theta] + H[ v_2 - \theta] \right] + D_v (v_2 - v_1).
\end{align*}
Thus, bees are either committed to foraging or uncommitted (can be recruited). This simplification is also relevant to phenomenon by which honey bees switch roles fluidly according to colony needs, so they may readily move between pure foraging and mixed foraging/waggling behaviors~\citep{johnson03,ribbands52,biesmeijer01}.
%
In this case, the colony can either be fully recruited long term ({\bf strong recruitment}) or recruitment may cease with the colony only partially recruited through the first patch ({\bf transient recruitment}). We can delineate these cases with their respective fixed points. Since Eq.~(\ref{discreteEqn}) is piecewise linear, there are four classes of fixed points: (i)~$v_1, v_2 > \theta$; (ii)~$v_1 > \theta > v_2$; (iii)~$v_1, v_2 < \theta$; and (iv)~$v_2 > \theta > v_1$. Type (iv) is impossible when $v_1(0) > v_2(0) = 0$ since when $v_1 = v_2$
\begin{align*}
\frac{d}{dt} (v_1 - v_2) = \beta [H(v_1- \theta) w_1 + H(v_2 - \theta) w_2 ] \geq 0.
\end{align*}
The other three types of fixed points are possible though: Type (i) implies that at equilibria:
\begin{align*}
\beta w_j = D_w (w_k - w_j), \ (j,k) \in \{ (1,2); (2,1) \}; \ \ \beta (w_1 + w_2) = D_v (v_1 - v_2), \ \  v_1 = v_2;
\end{align*}
which can be solved for $\bar{v}_1 = \bar{v}_2 = 0.5$ using fact that $v_1 + v_2 + w_1 + w_2 = 1$, the long term equilibrium associated with strong recruitment, which is marginally stable as indicated by the non-positive eigenvalues $\lambda = \{ 0 , -2D_v, -2D_w - \beta, \beta \}$. Type (ii) equilibria satisfy the fixed point equations
\begin{align*}
\beta w_1 = D_w (w_2 - w_1); \ \ w_1 = w_2; \ \ \beta w_1 = D_v (v_1 - v_2); \ \ v_1 = v_2;
\end{align*}
implying $w_1 = w_2 = 0$ so $v_1 = v_2 = 0.5$ which violates the inequality condition ($v_1 > \theta > v_2$) for this class of equilibria, making the intermediately recruited steady state impossible in the absence of foragers. Lastly, type (iii) equilibria satisfy $w_1 = w_2 = \bar{w} $ and $v_1 = v_2 = \bar{v} < \theta$, which along with the conservation condition implies the line of fixed points, $\bar{w} = 0.5 - \bar{v} > 0.5 - \theta$, corresponds to transient recruitment.

In the case of an approach to a type (iii) equilibrium, we can demonstrate that all bees are recruited via the first patch. Assume for contradiction that some recruitment occurs in the second patch. As mentioned previously, this implies $v_1 > v_2 > \theta$ for some time $t$. In this case, $v_1 > v_2 > \theta$ indefinitely thereafter, since $v_1 + v_2 > 2 \theta$, implying that at $v_2 = \theta$ we have $v_1 > \theta$ and $v_2'(t) = D_v ( v_1 - \theta) > 0$, which cannot be if a type (iii) equilibrium is approached.

Thus, colonies that begin in the type (i) ($v_1, v_2>\theta$) or (iii) ($v_1, v_2 < \theta$) regions remain there indefinitely, and there is a separatrix in the type (ii) region ($v_1> \theta > v_2$) delineating situations in which strong versus transient recruitment occurs long term. In this case, the evolution equations for the population in the type (ii) region are given by
\begin{align}
w_1' = - \beta w_1 + D_w (w_2 - w_1); \ \ w_2' = D_w (w_1 - w_2); \ \ v_1' = \beta w_1 + D_v (v_2 - v_1); \ \ v_2' = D_v (v_1 - v_2).  \label{type2eq} 
\end{align}
A steady state analysis is thus insufficient to separate population behavior when initial conditions begin in region (ii). As such, we will examine the dynamics of Eq.~(\ref{type2eq}) as it depends on model parameters. While we can derive full solutions to Eq.~(\ref{type2eq}) by hand, the implicit transcendental equations bounding different phases of recruitment are most tractable in the limiting cases we explore now: \\
\vspace{-3mm}

\noindent
{\bf Slow recruitment: $\beta/D_w \to 0$ and $\beta/D_v \to 0$.} In the limit of slow recruitment, wagglers recruit uncommitted bees much slower than bees move about the hive. As such, the patchiness of the hive does not matter as the colony becomes well mixed quickly. On fast timescales, $w_1, w_2 \to \tilde{w}$ (average of each other) and $v_1, v_2 \to \tilde{v}$. Combining the $w$ and $v$ equations, on slow timescales, we then have $\tilde{w}' = - \beta \tilde{w}$ and $\tilde{v}' = \beta \tilde{w} = \beta (0.5 - \tilde{v})$ by conservation. Recruitment will then cease if the limiting fraction of wagglers in patch 1 in this initial phase ($v_1 \to v_1(0)/2$) is below threshold $\theta$. Otherwise, if $v_1(0) > 2\theta$, recruitment continues till the whole colony is recruited. Thus, if recruitment is slow, the colony must rely on a sufficiently large initial set of wagglers to diffuse throughout the hive without falling subthreshold. \\
\vspace{-3mm}

\noindent
{\bf Fast recruitment: $\beta/D_w \to \infty$ and $\beta/ D_v \to \infty$.} If recruitment is much faster than the movement of the bees within the hive, then wagglers in patch 1 quickly recruit all bees there so $v_1 \to v_1(0) + w_1(0)$ initially. If $w_1(0) > v_1(0) > \theta$, recruitment will continue until the whole colony is recruited since diffusion of the recruiters will bring both patches above threshold. On the other hand, if $w_1(0) + v_1(0)< 2 \theta$, we can use the fact that on slow timescales $w_1 \approx 0$ and conservation to derive the slow subsystem
\begin{align*}
v_1' = D_w(1-v_1 - v_2) + D_v(v_2 - v_1); \ \ v_2' = D_v(v_1 - v_2).
\end{align*}
As the bees in patch 1 are quickly recruited, then at $0<t_1 \ll 1$, $v_1(t_1) \approx w_1(0) + v_1(0)$ and $v_2(t_1)=0$ which results in the following solutions to leading order
\begin{subequations} \begin{align} 
v_1(t) = & \frac{1}{2} + \frac{e^{-2 t D_v} \left( ( 1+v_1(0)) D_v-D_w\right)}{2 \left(2 D_v-D_w\right)} + \frac{e^{-t D_w} (1-v_1(0) ) (D_w-D_v)}{ 2 \left( 2 D_v-D_w \right)} \\
v_2(t) = & \frac{1}{2} - \frac{e^{-2 t D_v} \left( ( 1+v_1(0)) D_v-D_w\right)}{2 \left(2 D_v-D_w\right)} - \frac{e^{-t D_w} D_v(1-v_1(0) )}{ 2 \left( 2 D_v-D_w \right)} .		
\end{align}	\label{sol2pbetainf}
\end{subequations}
Checking whether or not strong recruitment occurs then depends on whether $v_1(t) < \theta$ at some point in Eq.~(\ref{sol2pbetainf}). If recruitment is sustained in the first patch, the fraction of wagglers in the second patch will continue to increase above threshold $\theta$, leading to strong recruitment.

To study whether the fraction of recruiters within the first patch is increasing or not, we differentiate $v_1(t)$ and determine bounds on when it is positive or negative. When $ v_1(0) \in [0,\frac{1}{2}]$, we find $v_1'(t) >0$ (so $v_1(t)$ increases monotonically) if $ D_w < D_v < \frac{D_w (1-v_1(0))}{v_1(0)}$, in which case recruitment will continue indefinitely. On the other hand, when $ D_v \geq \frac{D_w(1-v_1(0))}{ v_1(0)}$ $ \left( D_v < D_w \right)$, $v_1(t)$ is initially decreasing (increasing) and obtains a minimum (maximum). This critical point occurs where
\begin{align*}
v_1'(t^*) = & -\frac{D_v e^{-2 t D_v} \left( ( 1+v_1(0)) D_v-D_w\right)}{2 D_v-D_w}-\frac{D_w \left(1 - v_1(0) \right) e^{-t D_w} \left(D_w-D_v\right)}{ 2 \left(2 D_v-D_w\right)} = 0 \\
t^* = & \frac{\log ( \frac{(1- v_1(0)) D_w \left(D_v-D_w\right)}{2 D_v \left(( v_1(0)+1) D_v-D_w\right)} )}{D_w-2 D_v}.
\end{align*}
Thus, the critical point is 
\begin{align}   \label{v1min2patchbetainf}
v_1(t^*) = \frac{1}{2} + \frac{\left(D_w- (1+ v_1(0)) D_v \right)}{2 D_w} \left( \frac{ 2 D_v \left( (1 + v_1(0) ) D_v-D_w \right) }{D_w(1 - v_1(0))  (D_v -D_w) } \right)^{\frac{2 D_v}{D_w-2 D_v}}.
\end{align}	
If Eq.~(\ref{v1min2patchbetainf}) is less than the recruitment threshold ($v_1(t^*) < \theta$), recruitment terminates, but this will never occur when $\theta < \frac{1+v_1(0)}{4} $ since Eq.~(\ref{v1min2patchbetainf}) has a lower bound of $\frac{1+v_1(0)}{4}$. Note that for large values of recruitment thresholds, the critical waggler diffusion level $\hat{D}_v$ will depend linearly on the rate of uncommitted bee diffusion $\hat{D}_v = \alpha \cdot D_w$, following an argument similar to one provided in the slow uncommitted movement section next, based on the IVT.  \\

Lastly, we can consider individual limits of movement within the hive of both the wagglers and uncommitted bees. \\
\vspace{-3mm}

\noindent
{\bf Slow uncommitted movement: $D_w \to 0$.} This assumption simplifies the dynamics of the uncommitted population so there is no movement between patches: $w_1' = - \beta w_1$ and $w_2' = 0$ in the type (ii) region. Here, again, we can reduce the four-dimensional Eq.~(\ref{type2eq}) to a phase plane in the first time epoch using the fact that $w_2(t) = (1-v_1(0))/2$ so $w_1 = (1+v_1(0))/2 - v_1 - v_2$ and thus
\begin{align*}
v_1' = \beta \left[ \frac{1 + v_1(0)}{2} - v_1 - v_2 \right] + D_v(v_2 - v_1); \ \ v_2' = D_v (v_1 - v_2).
\end{align*}
We solve this system to obtain the following leading order equations
\begin{subequations} 	\begin{align}
v_1(t) =&  \frac{1+v_1(0)}{4} +  \frac{(D_v-\beta)(1-v_1(0))}{2(\beta - 2 D_v)} e^{- \beta  t} - \frac{4 D_v v_1(0) -\beta(1+v_1(0))}{4(\beta - 2 D_v)} e^{-2 D_v  t}, \\
v_2(t) =&  \frac{1+v_1(0)}{4}  + \frac{D_v(1-v_1(0))}{2(\beta - 2 D_v)} e^{- \beta  t} + \frac{4 D_v v_1(0) -\beta(1+v_1(0))}{4(\beta - 2 D_v)} e^{-2 D_v  t}. 
\end{align} \label{sol2patchDwzero}	\end{subequations}
As in the case of fast recruitment, continued recruitment in the first patch leads to strong recruitment of the entire hive. However, if there is a low initial fraction of recruiters $v_1(0)$ or if the diffusion of recruiters $D_v$ is rapid, the fraction of waggle dancing bees in the first patch decreases and recruitment may cease.

We can differentiate $v_1(t)$ in Eq.~(\ref{sol2patchDwzero}a) and find that it increases monotonically (obtains a maximum) when  $ \beta \leq D_v \leq \frac{\beta(1-v_1(0))}{2 v_1(0)} $ ($D_v < \beta$) if $v_1(0) \in [0,\frac{1}{3}]$, so recruitment continues indefinitely in this case. Alternatively, $v_1(t)$ obtains a minimum when $ D_v \geq \frac{\beta(1-v_1(0))}{2 v_1(0)}$ if $v_1(0) \in [0,\frac{1}{3}]$, or when $D_v > \beta$ if $v_1 \in (\frac{1}{3},1].$ To find the minimum, we solve for the time when
\begin{align*}
v_1'(t^*)  = & - \frac{ \beta (1 - v_1(0) ) (D_v - \beta ) e^{- \beta }}{2 (\beta - 2 D_v )} + \frac{D_v e^{-2 D_v t} (4 v_1(0) D_v  - \beta(1+v_1(0) ) )}{2 (\beta - 2 D_v  )} = 0\\
t^* = & \frac{\ln \left(\frac{ -4 D_v^2 v_1(0) + \beta D_v (1 + v_1(0)) }{\beta (v_1(0)-1)(D_v -\beta)} \right)}{ 2 D_v -\beta}.
\end{align*}
Thus, the minimum is
\begin{align} 	 \label{v1min2patchDwzero}
v_1(t^*) =& \frac{1 + v_1(0)}{4} +  \frac{\beta  (1 + v_1(0) ) - 4 D_v v_1(0)}{4 \beta }  \left(\frac{ - 4 D_v^2 v_1(0) + D_v \beta + \beta D_v v_1(0) }{\beta  (v_1(0)-1) (D_v-\beta )}\right)^{\frac{2 D_v}{\beta -2 D_v}} .		
\end{align}
If this minimum is less than the recruitment threshold fraction, recruitment terminates. We obtain the critical level of diffusion separating strong and transient recruitment by solving for this $D_v$ corresponding to where $v_1(t^*)= \theta$ in Eq.~(\ref{v1min2patchDwzero}).

The relationship between critical $D_v$ and recruitment rate $\beta$ is shown in Fig~\ref{fig2}c. We affirm the linear relationship between critical $D_v$ and recruitment rate $\beta$ as seen in Fig~\ref{fig2}c. Using the anstaz $D_v = \alpha \beta $, equation for critical diffusion level becomes
\begin{align*} 
f(\theta, v_1(0), \alpha) = \theta - \frac{1 + v_1(0)}{4} -  \frac{  1 + v_1(0)  - 4 \alpha v_1(0)}{4 }  \left(\frac{ \alpha (- 4 \alpha v_1(0) + 1 +v_1(0)) }{ (v_1(0)-1) (\alpha -1 )}\right)^{\frac{2 \alpha}{1 -2 \alpha}} .		
\end{align*}
The slope of the linear relationship between the recruitment rate $\beta$ and the critical value $D_v$ must satisfy $ \alpha \geq \frac{1-v_1(0)}{2 v_1(0)}$. Note that $f\left( \theta, v_1(0), \frac{1-v_1(0)}{2 v_1(0)} \right)  = \theta - v_1(0)$ and $\lim_{\alpha \to \infty} f(\theta, v_1(0), \alpha) = \theta - \frac{v_1(0)}{2}$ thus if $\theta < v_1(0) < 2\theta$ (which is our assumption), then due to the continuity of $f(\theta, v_1(0), \alpha) $, it has a root guaranteed by the Intermediate Value Theorem (IVT). \\

\noindent
{\bf Rapid uncommitted movement: $D_w \to \infty$.} In this limit, the uncommitted population equilibrates to be well mixed (so $w_1 = w_2 = \tilde{w}$), so conservation implies
\begin{align*}
v_1' = \beta \left[ \frac{1 - v_1 - v_2}{2} \right] + D_v(v_2 - v_1); \ \ v_2' = D_v (v_1 - v_2).
\end{align*}
which we solve to obtain the limiting evolutions of the waggler populations:
\begin{subequations} 	\begin{align}
v_1(t) =& \frac{1}{2} +  \frac{4 v_1(0)D_v-\beta}{2(4 D_v- \beta)} e^{- 2 D_v t} + \frac{(1- v_1(0)) (\beta - 2 D_v)} {4 D_v- \beta} e^{- \frac{\beta}{2} t}, \\
v_2(t) =& \frac{1}{2} -  \frac{4 v_1(0)D_v-\beta}{2(4 D_v- \beta)} e^{- 2 D_v t} - \frac{(1- v_1(0)) 2 D_v} {4 D_v- \beta} e^{- \frac{\beta}{2} t} .
\end{align}	\label{sol2patchDwinf}		\end{subequations}
As opposed to previous case, uncommitted bees from the second patch can diffuse to the first patch and are recruited before recruitment begins in the second patch. However, recruitment can terminate if the diffusion $D_v$ of wagglers is high or the initial fraction of recruiters $v_1(0)$ is too low. Using a similar approach to the cases before, we find $v_1(t)$ is increasing when $ D_v < \frac{\beta(1-v_1(0))}{2 v_1(0)}$.  Thus, the colony achieves strong recruitment. On the other hand, $v_1(t)$ attains a minimum when $ D_v \geq \frac{\beta(1-v_1(0))}{2 v_1(0)}$ if $v_1(0) \in [0,\frac{1}{2}]$, or when $D_v > \frac{\beta}{2}$ if $v_1 \in (\frac{1}{2},1]$, which occurs where
\begin{align*}
v_1'(t^*) = &  \frac{-D_v(4 v_1(0)D_v-\beta)}{4 D_v- \beta} e^{- 2 D_v t} + \frac{\beta (1- v_1(0)) (\beta - 2 D_v)} {2(4 D_v- \beta)} e^{- \frac{\beta}{2} t} = 0 \\ 
t^* = & \frac{2 \ln \left(-\frac{2 (4 v_1(0)  D^2-\beta  D )}{(v_1(0) -1) \beta  (2 D-\beta )}\right)}{4 D-\beta },
\end{align*}
implying the critical point occurs at
\begin{align} 	 \label{v1min2patchDwinf}
v_1(t^*) = \frac{1}{2} + \frac{ (\beta -4 v_1(0)  D_v)}{2 \beta } \left( \frac{ 2 D_v (\beta -4 v_1(0)  D_v) }{ (v_1(0) -1) \beta (2 D_v - \beta )} \right)^{\frac{4 D_v}{\beta -4 D_v}} .
\end{align}
If this minimum is less than threshold $v_1(t^*) < \theta$, recruitment eventually terminates. We characterize the regions of strong and transient recruitment using this inequality. The relationship between critical $\hat{D}_v$ and recruitment rate $\beta$ is shown in Fig~\ref{fig2}b as determined by Eq.~(\ref{v1min2patchDwinf}), and it is clear to see it is linear.

To show this, consider the ansatz $\hat{D}_v = \alpha \beta $, equation for critical diffusion level now becomes
\begin{align*} 
f(\theta, v_1(0), \alpha) = \theta - \frac{1}{2} - \frac{ (1 -4 v_1(0)  \alpha)}{2} \left( \frac{ 2  \alpha (1-4 v_1(0)   \alpha) }{ (v_1(0) -1) (2  \alpha - 1 )} \right)^{\frac{4  \alpha}{1 -4  \alpha}} .	
\end{align*}
For initial recruiters fraction in the interval $[0, 1/2]$, $v_1(t)$ obtains a minimum when $ \alpha \geq \frac{1-v_1(0)}{2 v_1(0)}$. Noting then that $f\left( \theta, v_1(0), \frac{1-v_1(0)}{2 v_1(0)} \right) = \theta - v_1(0)$ and $\lim_{\alpha \to \infty} f\left( \theta, v_1(0), \alpha \right) = \theta - \frac{v_1(0)}{2}$, we see that if $\theta < v_1(0) < 2 \theta$ (which is our assumption), $f(\theta, v_1(0), \alpha) $ has a root guaranteed by the IVT. \\

\noindent
{\bf Slow waggler movement: $D_v \to 0$.} In this limit, wagglers from patch 1 never reach patch 2, so $v_2 = 0$ indefinitely. In this case, clearly $v_1$ is always increasing since $v_1' = \beta w_1 > 0$, so we will always have strong recruitment. All uncommitted bees will be recruited via patch 1, as determined by the phase plane dynamics:
\begin{align*}
v_1' = \beta w_1; \ \ w_1' = - \beta w_1 + D_w ( 1- v_1 - 2w_1).
\end{align*}
This system obtains its fixed point $v_1 = 1$ and $w_1 = 0$ with eigenvalues $\lambda = \frac{1}{2} \left(- \beta - 2 D_w \pm \sqrt{\beta ^2+4 D_w^2} \right)$, which are both clearly negative and the fixed point is stable. \\

\noindent
{\bf Rapid waggler movement: $D_v \to \infty$.} Lastly, wagglers may rapidly equilibrate across the hive. In this case, to determine whether strong or transient recruitment occurs, we simply check if $v_1(0) > 2 \theta$ or not. If $v_1(0) < 2 \theta$, then $v_1, v_2 \to \tilde{v} < \theta$ rapidly and the colony falls into the type (iii) region and recruitment (and all dynamics) ceases. Otherwise, $v_1, v_2 \to \tilde{v} > \theta$, which together with conservation means the dynamics is described by the scalar equation for $w_1 = w_2 = w$, $w' = - \beta w$, so $w(t) = \frac{1-v_1(0)}{2} \e^{- \beta t}$ and $\tilde{v}(t) = 0.5 + 0.5 (v_1(0) - 1) \e^{- \beta t}$. This is consistent with our observation that the parameter space regions in which we expect a tradeoff in varying the diffusion of the wagglers are those for which $\theta < v_1(0) < 2 \theta$ in the two patch model. 

In total, our analysis shows that the initial fraction of wagglers in patch 1, $v_1(0)$, must be sufficiently large, their rate of diffusion $D_v$ must be sufficiently small, and the recruitment rate $\beta$ must be large enough for recruitment to persist until the whole hive is recruited. The above asymptotic results reveal clear trends in the relationship between these critical parameters that divide strong and transient recruitment. In particular, as $\beta$ increases, the critical $D_v$ increases linearly. Increasing the recruitment rate thus affords the colony the ability to spread information throughout the hive faster without the recruiters losing their critical capacity to influence the uncommitted. Similarly, we identified a linear relationship between $D_w$ and the critical diffusion rate $D_v$, since uncommitted movement can also facilitate recruitment. Moving forward, we now identify trends between these extremes, and show again that a colony must balance a trade off between rapidly propagating recruitment into the second patch, while preventing too rapid a dispersion of wagglers.

\subsubsection{Critical diffusion level of waggling bees in a hive with two compartments}
\begin{figure}[t]
\begin{center}    \includegraphics[width=\textwidth]{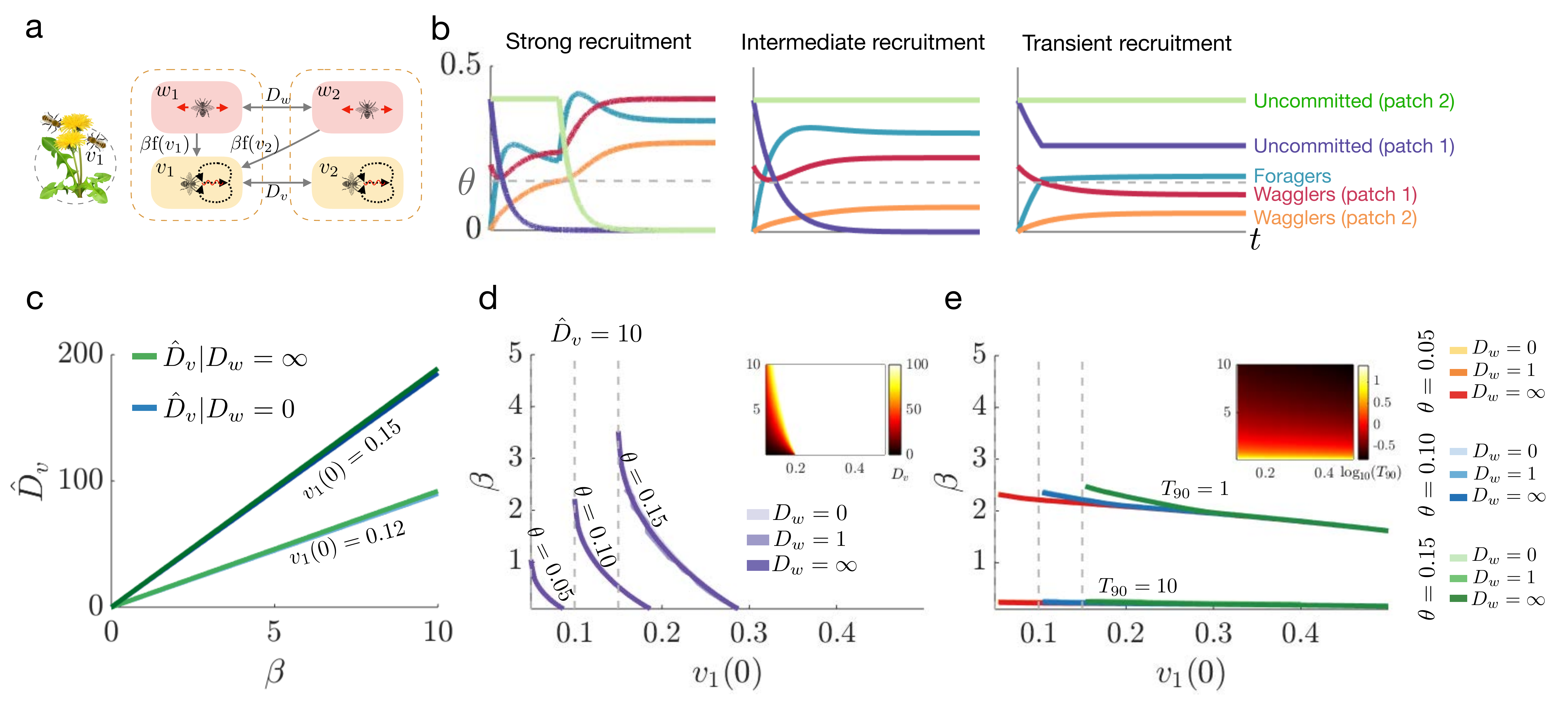}	\end{center} 		
\caption{ {\bf a.} Two patch hive model schematic without foragers. {\bf b.} Numerical simulations of Eq.~(\ref{discreteEqn}) resulting in three different recruitment scenarios: {\bf strong recruitment} (the fraction of recruiters increases in both patches and recruitment commences in second patch), {\bf intermediate recruitment} (the fraction of recruiters might decreases in first patch but does not fall below the minimum required for recruitment but recruitment does not commence in second patch), 
{\bf transient recruitment} (the fraction of recruiters decreases in the first patch  and falls below the minimum required for recruitment and recruitment ceases in both patches).  {\bf c.} The critical level of diffusion of wagglers $\hat{D}_v$ which causes recruitment to cease for different initial values of recruiters $v_1(0)$ and $\theta = 0.1$ in the limits of movement of uncommitted bees. {\bf d.} Level curves where the critical diffusion coefficient for waggling bees is $\hat{D}_v = 10$. Along and to the left of these curves, if $D_v \geq 10$, strong recruitment ceases.  The threshold of waggling bees required for recruitment varies as $\theta = 0.05, 0.10, 0.15$. The diffusion rate $\hat{D}_v$ at which recruitment ceases for $\theta = 0.10$ is shown in the inset. {\bf e.} Level curves for time to recruit $90 \% $ hive population at the optimal $D_v$ (in log scale), which was the value just below the critical diffusion level. The inset shows the time to recruit $90 \%$ of the hive population when $\theta = 0.10$ and $D_w = 0$ in log scale.}
\label{fig2}
\end{figure}

The speed at which a colony can propagate information about potential foraging sites and recruit the hive strongly depends on the fraction of initial recruits $v_1(0)$ waggle dancing for the foraging sites. A high fraction of initial recruits facilitates rapid information transfer via the diffusion of waggle dancing bees in the hive. In contrast, if the initial fraction of recruiters is too low, strong diffusion can spread the population of wagglers too thin. This suggests that there is an optimal level of movement of waggle dancing bees that allows the colony to quickly recruit the hive without spreading the wagglers below the recruitment threshold. We determine the level of diffusion that separates strong from transient recruitment across multiple parameter regimes. Then, assuming recruiters diffuse at a rate close to this critical level $\hat{D_v}$, we calculate the shortest time to recruit $90 \%$ of bees in the hive, $T_{90}$, for the hive with two patches. In general, we find a rate of diffusion just below $\hat{D_v}$ facilitates rapid transfer of information in the hive while ensuring the eventual recruitment of all uncommitted bees in a two patch and three patch hives with a central entrance.

We did find that as the number of patches in the hive increases, the level of diffusion close to the critical level $\hat{D_v}$ is not optimal for rapid information transfer, since it takes longer for the colony to reach a super-threshold waggler density across all patches. Instead, the optimal level $D_v^{\rm opt}$ for maximizing the rate of 90\% recruitment is notably lower than the critical value $\hat{D}_v$. This lower level of diffusion leads wagglers to move slowly in the hive, increasing waggler density in patches adjacent to the entrance, and allowing recruitment to commence in more than one patch faster than when wagglers spread rapidly across the hive. Thus, for hives with multiple peripheral patches, we calculated $T_{90}$ by optimizing time to recruit $90 \%$ of the hive across all possible levels $D_v$ of waggler diffusion, as we show subsequently.

In a hive with just two compartments, the critical level of diffusion of wagglers $\hat{D}_v$ is insensitive to changes in the diffusion rate $D_w$ of uncommitted bees (Fig~\ref{fig2}d). Diffusion of uncommitted bees is not a significant driver of recruitment dynamics in a two patch hive when the uncommitted population is initially spread evenly in the hive. If $D_w$ is large, the population of uncommitted bees will stay balanced throughout the recruitment process. If $D_w$ is small, a substantial fraction of uncommitted bees remains in the second patch until the waggler population there becomes superthreshold, at which point both patches will be recruited at an equal rate leading to larger fractions of recruitment in the second patch, compensating for the lack of initial recruitment there. This is why the shortest time to recruit $90 \%$ of the hive ($T_{90}$) does not significantly depend on the movement of uncommitted bees in the hive (Fig.~\ref{fig2}e).  Note also that as the rate of recruitment $\beta$ is decreased, the critical rate $\hat{D}_v$ decreases, since the waggler population must not be spread too thin too quickly.

In the following section, we show that the movement of uncommitted bees expedites recruitment when the number of patches in the hive is higher as there is less crowding around the waggle dancing bees when bees are spread out in higher number of patches.

\subsubsection{Recruitment dynamics in three hive patch model}

\begin{figure}[t]
\begin{center}    \includegraphics[width=\textwidth]{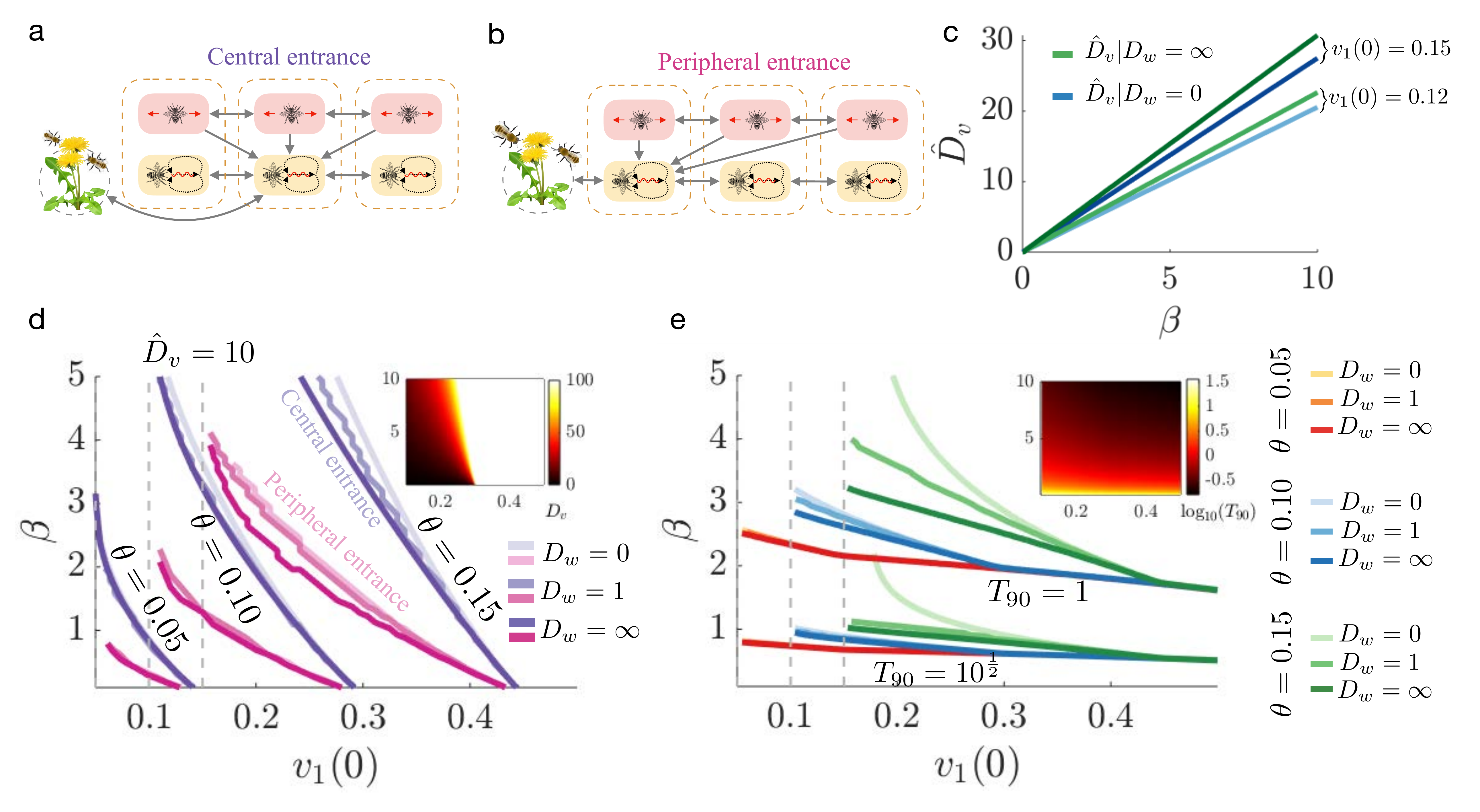}	\end{center} 	
\caption{ Three patch hive model. {\bf a.} Hive with entrance in the center patch. {\bf b.} Entrance in the peripheral patch. {\bf c.} The critical level of diffusion of wagglers $\hat{D}_v$ halting recruitment for different initial values of recruiters $v_1(0)$ and $\theta = 0.1$ when diffusion of uncommitted bees is slow $D_w = 0$ (blue) and diffusion of uncommitted bees is rapid $D_w = \infty$ (green). {\bf d.} Level curve for $\hat{D}_v = 10$ when the proportion of waggling bees required for recruitment varies as $\theta = 0.05, 0.10, 0.15$. Purple curve represents central entrance and pink curve represents peripheral entrance. The diffusion rate $\hat{D}_v$ at which recruitment ceases for $\theta = 0.10$ is shown in the inset; {\bf e.} Level curves for time to recruit $90 \% $ hive population at optimal $\hat{D}_v$ (in natural log scale).  We choose diffusion of waggling bees just below the critical diffusion level $\hat{D}_v$ to be optimal for the speedy recruitment of the hive in the central entrance hive but compute $T_{90}$ in peripheral case through numerical optimization across $D_v$.  The figure shown is for the central entrance case, however the peripheral case looks nearly the same. The inset shows the time to recruit $90 \%$ of the hive population when $D_w = 0$ and $\theta = 0.1$ in natural log scale.}
\label{fig3}
\end{figure}

In a hive with three compartments, movement becomes more import to spread information throughout the spatially-distributed hive. The population of wagglers must tune their movement to effectively reach and recruit uncommitted bees throughout the hive. We consider scenarios in which waggle dancing bees can enter the hive through the compartment at the center (central entrance: Fig~\ref{fig3}a) or at the either end (peripheral entrance: Fig~\ref{fig3}b). As in the two compartment hive, the movement of waggle dancing bees determines the progression of recruitment in the hive. Rapid movement of wagglers allows the colony to swiftly recruit all members for a timely exploitation of resources, but at the risk of losing recruiters in the initial compartment, halting recruitment. Given the hive is distributed across more patches, there is an increased risk of spreading too thin, and the entire recruitment process ceasing.

In the absence of the foraging class, the fraction of committed and uncommitted bees in the central entrance hive with three compartments evolve as follow
\begin{subequations} 	\begin{align}
\frac{\partial w_i}{\partial t}  =&  -\beta H[v_i - \theta] w_i + \begin{cases}  D_w(w_{j \pm 1}-w_j), \quad & j=1,3 \\ D_w(w_{3}-2 w_2 + w_{1}), & j = 2,	\end{cases} \\
\frac{\partial v_i}{\partial t}  =&  \begin{cases}  D_v(v_{2}-v_j), \quad & j=1,3, \\   \beta \sum_{i = 1}^{3} H[ v_i-\theta] w_i  + D_v(v_{3}-2 v_2 + v_{1}), & j=2. \end{cases} 
\end{align} \label{eq3patch}	\end{subequations}
As in the two patch case, we can determine the relative role of recruitment and bee movement in the dynamics of the collective recruitment process in Eq.~(\ref{eq3patch}) by studying its asymptotic limits in the case of central entrance hive: \\

\noindent
{\bf Fast recruitment: $\beta/D_w \to \infty$ and $\beta/D_v \to \infty$.} Wagglers in patch 2 are quickly recruited so $v_2 \to v_2(0) + w_2(0)$. If $v_2(0) + w_2(0) < 3\theta$, we can use that $w_2 \approx 0$, symmetry about the center patch ($v_1(t) = v_3(t)$), and conservation to derive the slow subsystem
\begin{align*}
v_1' =   D_v(v_2 - v_1), \ \
v_2' = D_w(1- v_2 -2 v_1) + 2 D_v(v_1 -v_2),
\end{align*}
which we solve to obtain in this limit:
\begin{subequations} 	\begin{align*}
v_1(t) =& v_3(t) = \frac{1}{3}  + \frac{2 ( 1 - v_2(0)) D_v }{3 (D_w-3 D_v)} e^{-t D_w} + \frac{D_w - (2 v_2(0) +1) D_v}{3 (3D_v-D_w)} e^{-3 t D_v},  \\
v_2(t) = &  \frac{1}{3}  - \frac{2 (1 - v_2(0)) (D_w-D_v )}{3 (D_w-3 D_v)}e^{-t D_w}  + \frac{ 2(D_v(2 v_2(0) + 1)  - D_w) }{3 (3 D_v - D_w )} e^{-3 t D_v},
\end{align*}	\end{subequations}
so that $v_2(t)$ attains a minimum where
\begin{subequations} 	\begin{align*}
v'_2(t^*) =& \frac{2 D_w (1 - v_2(0)) (D_w-D_v )}{3 (D_w-3 D_v)}e^{-t D_w}  - \frac{ 2 D_v (D_v(2 v_2(0) + 1)  - D_w) }{ 3 D_v - D_w } e^{-3 t D_v}  =0, \\
t^* =& \frac{\log \left(\frac{- v_2(0) D_v D_w+v_2(0) D_w^2+D_v D_w-D_w^2}{3 D_v \left(2 v_2(0) D_v+D_v-D_w\right)}\right)}{D_w-3 D_v}.
\end{align*}	\end{subequations}
The critical point is 
\begin{align}  \label{v1min3patchbetainf}
v_2(t^*) = \frac{1}{3}  + \frac{2 \left(D_w - D_v(2 v_2(0)+1) \right)}{3 D_w} \left(-\frac{3 D_v \left((2 v_2(0)+1) D_v-D_w\right)}{(v_2(0) -1) D_w \left(D_v-D_w\right)}\right)^{\frac{3 D_v}{D_w-3 D_v}}.
\end{align}
If this critical value is less than the recruitment threshold, recruitment will eventually terminate. However, this is not achieved for any level of waggler diffusion if the recruitment threshold $\theta < \frac{2 v_2(0)+1}{9}$ since in this case $\lim_{D_v \to \infty} v_2(t^*)=\frac{2 v_2(0)+1}{9}$, meaning strong recruitment will always occur if this inequality is satisfied. \\

\noindent
{\bf Slow uncommitted movement: $D_w \to 0$.} This simplifies the dynamics of the uncommitted population as there is no movement of uncommitted bees between the patches: $w_1' = w_3' = 0$ initially (before the waggler population in those patches becomes superthreshold) and $w_2' = \beta w_2$. Here, we can reduce the six dimensional Eq.~(\ref{eq3patch}) to two dimensions using the fact that $w_1(t) = w_3(t) = \frac{1-2v_1(0)}{3}$ so $w_2 = \frac{1+2v_1(0)}{3} - 2 v_1 - v_2$ and thus
\begin{align*}
v_1' =   D_v(v_2 - v_1); \ \
v_2' =  \beta \left( \frac{1+2v_1(0)}{3} - 2 v_1 - v_2 \right) + 2 D_v(v_1 -v_2),
\end{align*}
which we solve to obtain
\begin{subequations} 	\begin{align*}
v_1(t) =& v_3(t) =  \frac{1+ 2 v_2(0)}{9} + \frac{9 v_2(0)D_v - \beta - 2v_2(0) \beta }{3(\beta - 3D_v)} e^{-3 D_v  t} + \frac{ D_v(1- v_2(0))}{3(\beta - 3D_v)} e^{-\beta t} \\
v_2(t) =&  \frac{1+ 2 v_2(0)}{9}  -  \frac{2(9 v_2(0)D_v - \beta - 2v_2(0) \beta)}{3(\beta - 3D_v)} e^{-3 D_v  t} + \frac{ (D_v -\beta )(1- v_2(0))}{3(\beta - 3D_v)} e^{-\beta t},
\end{align*}	\end{subequations}
initially, and we find that $v_2(t)$ attains a minimum where
\begin{subequations} 	\begin{align*}
v'_2(t^*) =&   \frac{2 D_v (9 v_2(0)D_v - \beta - 2v_2(0) \beta)}{\beta - 3D_v} e^{-3 D_v  t} - \frac{\beta (D_v -\beta )(1- v_2(0))}{3(\beta - 3D_v)} e^{-\beta t}    =0 \\
t^* =& \frac{\log \left(-\frac{2 \left(9 v_2(0) D_v^2-2 v_2(0) \beta  D_v-\beta  D_v\right)}{(v_2(0)-1) \beta (D_v-\beta )}\right)}{3 D_v-\beta }.
\end{align*}	\end{subequations}
The critical point is 
\begin{align}  \label{v1min3patchDwzero}
v_2(t^*) = \frac{1}{9} (2 v_2(0) +1) + \frac{(1-v_2(0) ) (\beta -D_v)}{9 D_v} \left(\frac{2 D_v (2 v_2(0)  \beta -9 v_2(0) D_v + \beta )}{(v_2(0) -1) \beta  (D_v - \beta )}\right)^{\frac{\beta }{\beta -3 D_v}}.
\end{align}	
The relationship between critical diffusion of wagglers $\hat{D}_v$ and recruitment rate $\beta$ as determined by Eq.~(\ref{v1min3patchDwzero}) is shown in Fig.~\ref{fig3}c. Again, we could construct and argument to show $\hat{D}_v$ depends linearly on $\beta$. \\

\noindent
{\bf Rapid uncommitted movement: $D_w \to \infty$.} In this limit, the uncommitted population equilibrates to be well mixed (so $w_1 = w_2 = w_3 = \bar{w}$), thus conservation implies
\begin{align*}
v_1' =   D_v(v_2 - v_1); \ \
v_2' =  \beta \left( \frac{1 - v_1 - v_2 - v_3}{3} \right) + D_v(v_1 -2v_2 + v_3);  \ \
v_3' =   D_v(v_2 - v_3); 
\end{align*}
with solutions
\begin{subequations} 	\begin{align*}
v_1(t) =& v_3(t) =  \frac{1}{3} + \frac{3 (1 - v_2(0)) D_v }{\beta -9 D_v} e^{-\frac{\beta  t}{3}} - \frac{(\beta -9 v_2(0) D_v)}{3 (\beta -9 D_v)} e^{-3 t D_v},  \\
v_2(t) =& \frac{1}{3} -  \frac{(1-v_2(0)) (\beta -3 D_v)}{\beta -9 D_v} e^{-\frac{\beta  t}{3}} + \frac{2 (\beta -9 v_2(0) D_v )}{3 (\beta -9 D_v)} e^{-3 t D_v},
\end{align*}	\end{subequations}
so $v_2(t)$ attains a minimum where
\begin{subequations} 	\begin{align*}
v'_2(t^*) =&  \frac{\beta (1-v_2(0)) (\beta -3 D_v)}{3(\beta -9 D_v)} e^{-\frac{\beta  t}{3}} - \frac{2 D_v (\beta -9 v_2(0) D_v )}{\beta -9 D_v} e^{-3 t D_v}   =0, \\
t^* =& \frac{3 \log \left(-\frac{6 \left( 9 v_2(0) D_v^2-\beta D_v \right)}{(v_2(0) -1) \beta  (3 D_v - \beta)}\right)}{9 D_v - \beta }.
\end{align*}	\end{subequations}
The critical point is 
\begin{align}	 \label{v1min3patchDwinf}
v_2(t^*) = \frac{1}{3} + \frac{ 2(\beta -9 v_2(0) D_v)}{3 \beta} \left(\frac{6 D_v (\beta -9 v_2(0) D_v)}{(v_2(0) -1) \beta  (3 D_v-\beta )}\right)^{\frac{9 D_v}{\beta -9 D_v}}
\end{align}
so $\hat{D}_v$ occurs where $v_2(t^*)= \theta$ in Eq.~(\ref{v1min3patchDwinf}), which again depends linearly on $\beta$.

We calculate $\hat{D_v}$ for the peripheral entrance hive and intermediate values of $D_v$ in central entrance hive using numerical simulations. The critical level of diffusion of wagglers $\hat{D}_v$ in these cases is lower compared to the two patch case as bees have more space to spread out and  can more easily drop their patch populations below threshold, so there is a higher premium on maintaining a superthreshold population in the entrance patch. This effect is more drastic in the central entrance hive (Fig~\ref{fig3}d). In this case, bees diffuse in both directions from the entrance (`leaking' at twice the rate from the center patch), and the critical level of diffusion at which number of waggle dancing bees goes below the recruitment  threshold is lower, since the fraction of recruiters in the entrance patch decreases more rapidly.   

Adding a patch to the model slows recruitment since bees must now recruit from more patches and traverse  longer distances when the entrance is peripheral (Fig~\ref{fig3}d). Uncommitted bees in the hive are spread out further and wagglers need to be above threshold in each compartment to sustain strong recruitment. The movement of uncommitted bees facilitates the recruitment of the entire hive more so than in the two patch case, since there is more space to be covered (Fig~\ref{fig3}d).

If the colony tunes their diffusion rates with respect to the arrangement of the hive (central vs. peripheral entrance), there is little to no difference between the time to recruit $90 \%$ of the hive ($T_{90}$) for the central entrance case and peripheral entrance case (Fig.~\ref{fig3}e). This suggests that even though it may be easier for recruitment to fail in the central entrance case, the hive-wide rate of recruitment for an optimal tuned colony does not suffer in comparison to the hive with a peripheral entrance.

\subsubsection{Critical diffusivity $\hat{D_v}$ decreases with the number of patch compartments}

The movement of uncommitted bees  becomes more essential to the recruitment process as the number of hive compartments increases, since wagglers have more ground to cover to communicate recruitment signals. There is also a higher premium for sustaining superthreshold fractions of recruiters in each patch as number of patches grow which becomes unachievable at some patch number. In these cases, the only way for the entire hive to be recruited is through the movement of uncommitted bees to the compartments closer to the entrance with more recruiters, and having the recruiters stay confined primarily close to the entrance to maintain a superthreshold fraction, localizing the patch regions where recruitment occurs.

\begin{figure}[t!]
\begin{center}    \includegraphics[width=15 cm]{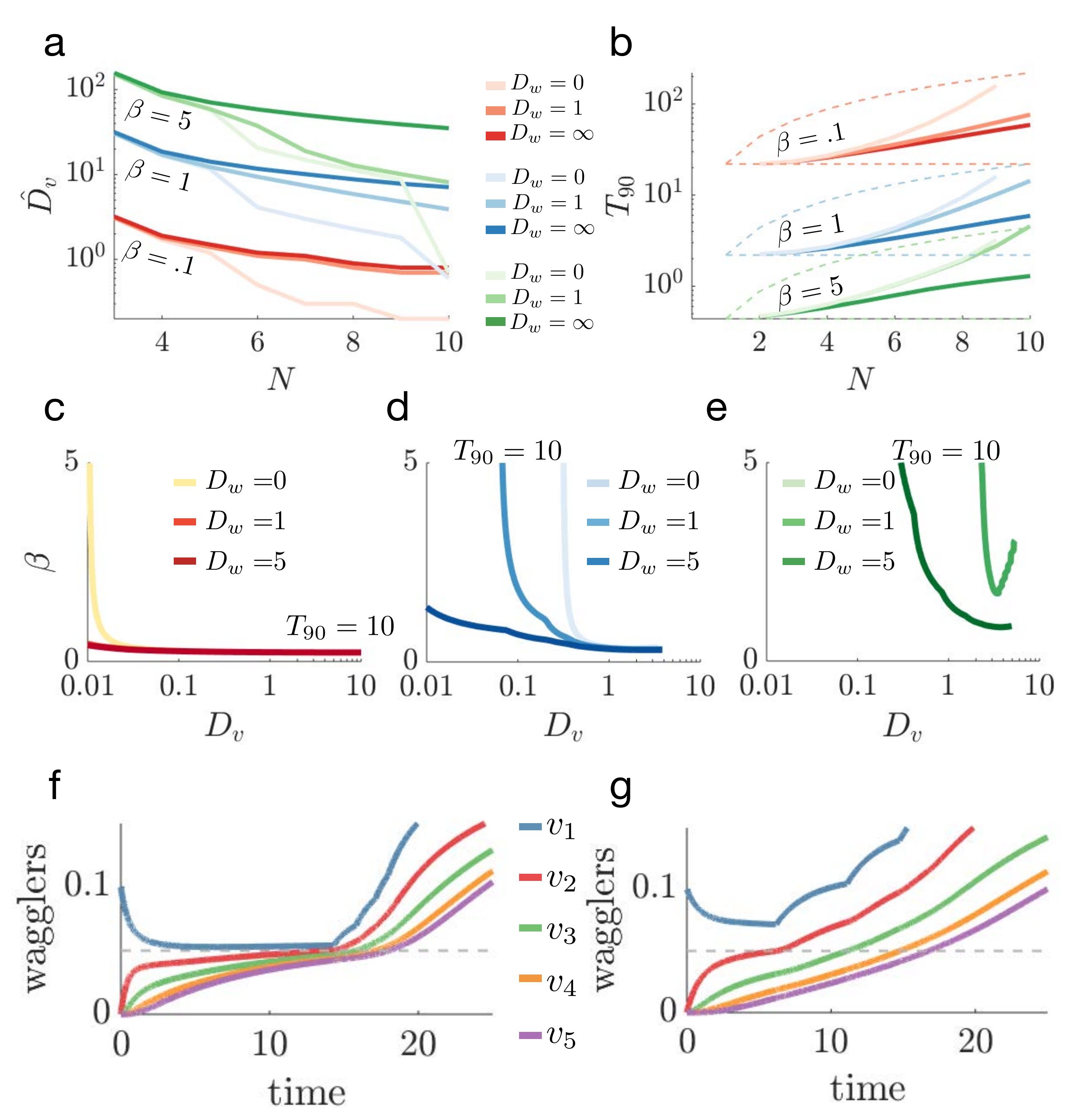}	\end{center} 	
\caption{  {\bf a.} Critical diffusivity level of waggling bees $\hat{D_v}$ at which value recruitment eventually ceases as a function of $N$, the number of patches in the hive. {\bf b.} Fastest time to recruit $90 \% $ hive population at critical $D_v$. For both figures, $\theta = 0.05$ and $v_1(0)=0.1$. Dashed lines form lower bounds for $T_{90}$ when $D_v, D_w = \infty$ (lower dashed line) and $D_v = 0, D_w = \infty$ (upper dashed curve). {\bf c,d,e.} Level curves for fastest time to recruit $90 \% $ hive population  $T_{90}$ for a hive with: {\bf c.} Two compartments {\bf d.} Five compartments; {\bf e.} Ten compartments {\bf f.}~Time series of the recruitment dynamics showing how waggler density across different patches evolves in a hive with five compartments: {\bf f.} $D_v = 1$ {\bf g.} $D_v = 0.48$. }
\label{fig4}
\end{figure}

Diffusion of uncommitted allows distant uncommitted bees to move closer to the entrance as those initially close to the entrance are recruited. As bees are recruited, the population of recruiters increases as well. Recruiters can diffuse without their populations in each patch dropping below the recruitment threshold.
If uncommitted bees diffuse slowly (low $D_w$), the critical level of recruiter diffusion $\hat{D_v}$ decreases rapidly with an increase in the number of compartments $N$, since a superthreshold fraction must be maintained with fewer recruited bees per patch. However, when uncommitted bees diffuse more rapidly (higher $D_w$), the critical level of diffusion $\hat{D}_v$ does not decrease as dramatically as more compartments are added (Fig~\ref{fig4}a), since more bees will be recruited due to the movement of uncommitted bees. 

Movement of uncommitted bees significantly reduces time to recruit $90 \%$ of the hive in a hive with large number of patches (Fig~\ref{fig4}b). Furthermore, we find $T_{90}$ does not decrease monotonically with the diffusion of wagglers even in the range of diffusion leading to super-threshold wagglers' density (Fig~\ref{fig4}e). In larger hives, large waggler diffusion delays the time taken for waggler population to be above threshold in peripheral patches thus prolonging time to recruit $90 \%$ of the hive (Fig.~\ref{fig4}f,g). It is more advantageous for a colony to quickly reach super-threshold waggler populations in the patches neighboring the entrance than to rapidly diffuse and equilibrate the population across hive, resulting in longer time for recruitment to commence in more than one hive. As number of patches in a hive increases, a balance of movement of uncommitted and committed bees is required to fully recruit the hive. 

We expect the time $T_{90}$ to recruit $90 \%$ of the hive to increase with the number of compartments in the hive and decrease with the increase in recruitment rate. Indeed, $T_{90}$ grows exponentially with $N$ when the population of uncommitted bees diffuses rapidly and supraexponentially when uncommitted bees diffuse slowly (Fig~\ref{fig4}b).

As limiting estimates of the curves for $T_{90}$ we have computed numerically, we can compute the time to recruit $90 \%$ directly in two scenarios: well mixed hive ($D_v , D_w \to \infty$) and stationary wagglers $D_v = 0, D_w \to \infty$. The numerically computed $T_{90}$ mostly fall in between these two values (Fig.~\ref{fig4}b).  The single patch hive or the well mixed hive forms lower bound for our model.  

\noindent {\bf Well mixed hive.} In a well mixed hive ($D_v , D_w \to \infty$), the number of uncommitted and committed bees in each compartment of a hive is equalized $w_i = \bar{w}$ and $v_i = \bar{v}$ which evolves as
\begin{align}
\bar{w}' = - \beta \bar{w},		\ \	\bar{v}' = \beta \bar{w},
\end{align}  
so $\bar{w}(t) = \frac{e^{-\beta t}}{N} \left(1-v_1(0) \right)$ and $\bar{v}(t) = \frac{1}{N} - \frac{e^{-\beta t}}{N} \left(1-v_1(0) \right)$, as long as $\theta< \frac{v_1(0)}{N}$. This is equivalent to the dynamics of a single patch hive with $1/N$ the fraction in the single patch. Assuming $\theta < \frac{v_1(0)}{N}$, recruitment continues until the hive is recruited, requiring a time to recruit $90 \%$ of the hive $T_{90} =- \frac{1}{\beta} \log \left(\frac{0.1}{1-v_1(0)} \right).$

\noindent
{\bf Stationary wagglers.} Next, assume no movement of wagglers ($D_v = 0$). For larger number of patches $N$, recruitment ceases for any level of diffusion of wagglers $D_v$ due to overspreading of wagglers across the hive, so the only way such a hive can obtain strong recruitment is if wagglers do not diffuse and recruitment occurs only in the first patch through diffusion of uncommitted bees. Thus, for the hives with a higher number of patches, $T_{90}$ is lower bounded by the case where $D_v = 0$ and $D_w = \infty$. Here, the fractions of uncommitted bees in each hive patch equalizes, $w_i = \bar{w}$ and $v_i = 0 \ (i \neq 1)$, and the fraction of wagglers in patch $1$ evolves as 
\begin{align}
v_1' =  \frac{\beta}{N}(1-v_1).
\end{align}  
Solving this and using conservation, $v_1(t)=  1- e^{\frac{-\beta t}{N}} \left(1-v_1(0) \right)$ and $\bar{w}(t) =e^{\frac{-\beta t}{N}} \left( \frac{1-v_1(0)}{N} \right)$. Again, we can directly compute $T_{90} =- \frac{N}{\beta} \log \left(\frac{0.1}{1-v_1(0)} \right).$ We plot both of these bounds for $T_{90}$ in Fig.~\ref{fig4}b along with the calculated values. The lower dashed line is $T_{90}$ corresponding to the well mixed assumption and the upper dashed curve corresponds to the stationary waggler case where $D_v = 0$ and $D_w = \infty$. For $N \leq 10$, the fastest time to recruit $90 \%$ of the population by optimizing the movement of wagglers falls between these two bounds. 

\subsection{Thresholded recruitment facilitates efficient collective signal detection}

\begin{figure}[t!]
\begin{center}    \includegraphics[width=15cm]{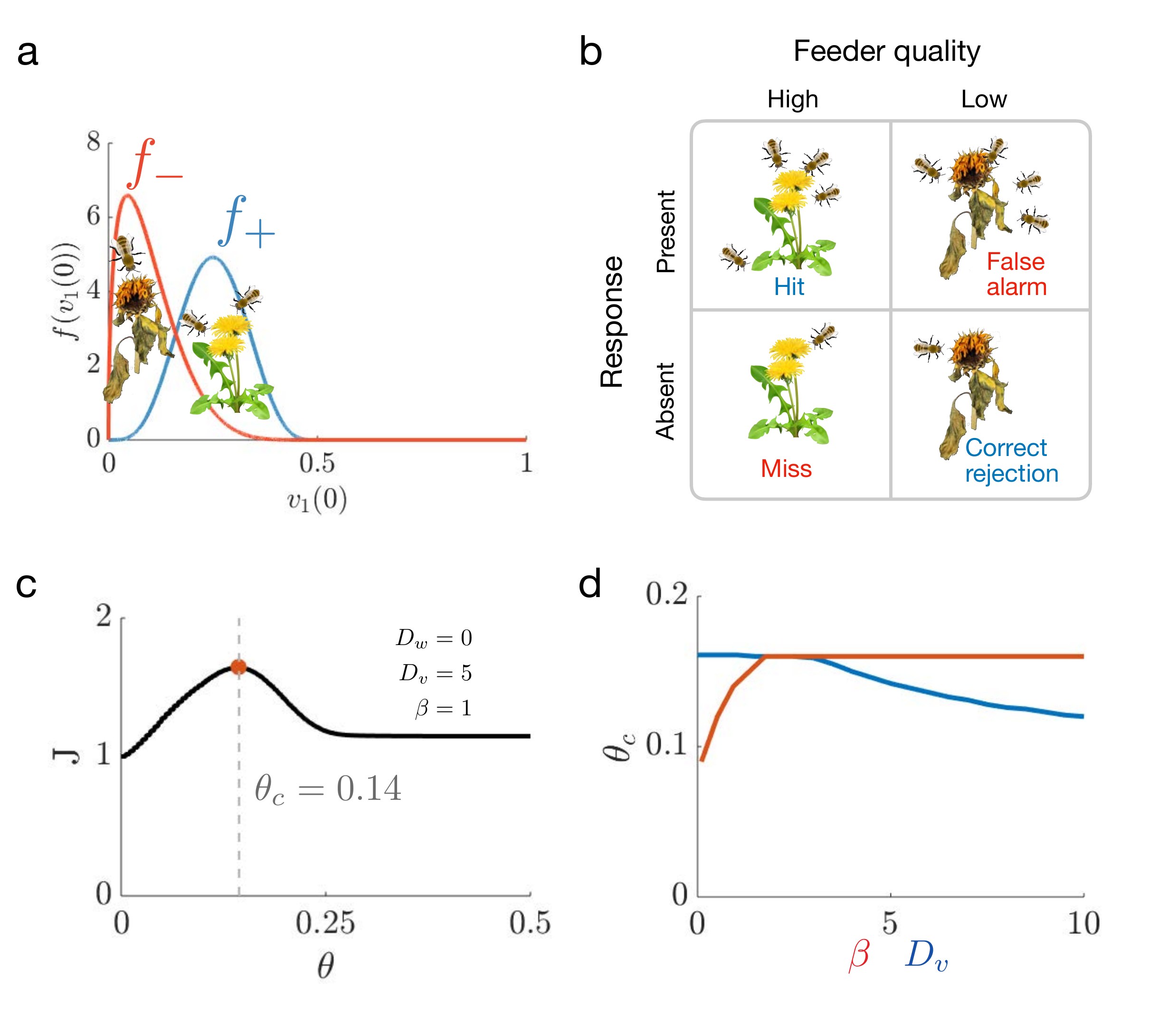}	\end{center} 	
\caption{  {\bf a.}~Probability density function for the fraction of scouting bees becoming recruiters after visiting a high yielding foraging source $f_{+}$ or a low yielding foraging source $f_{-}$. {\bf b.} Response matrix depicting all possible signal and response combination. Blue text indicates correct decisions and red text indicates incorrect decisions. {\bf c.} Objective function \emph{J}, the sum of the mean turn out of recruiters in the favorable state plus the mean of the fraction remaining uncommitted in the unfavorable state varies with thresholding value for $\beta = 1, D_v = 5, D_w = 0$.  {\bf d.} The threshold $\theta_c$ maximizing  $J$ varies with model parameters $\beta$, $D_v$ and $D_w$.}
\label{fig5}
\end{figure}

Though the threshold on recruitment can prevent and slow the initiation of colony-wide foraging, it can also prevent the colony from erroneously departing for low-quality food sources. To examine this tension between recruitment to high-quality foraging sites and prevention of departure to low-quality sites, we will examine a colony-wide thresholding mechanism that emerges to divide these two scenarios. In a simple hive with two compartments, using the following system of equations we can map the initial fraction of waggle dancing bees, $v_1(0)$ to the long term limit of committed bees, $v_s = \lim_{t \to \infty} v_1(t) + v_2(t)$. We define this mapping as the function $v_s \equiv g(v_1(0); \theta, \beta, D_w, D_v)$, where:
\begin{subequations} \begin{align}
\dot{w}_1 &= - \beta H[v_1 - \theta] w_1 + D_w (w_2 - w_1), \\
\dot{w}_2 & = - \beta H[v_2 - \theta] w_2 + D_w (w_1 - w_2), \\
\dot{v}_1 &= \beta (H[v_1 - \theta]w_1 + H[v_2 - \theta]w_2)  + D_v(v_2 - v_1), \\
\dot{v}_2 &= D_v(v_1 - v_2),
\end{align} \label{eq2patch}
\end{subequations} 
If recruitment commences in the second patch (strong recruitment cycle), $v_s = 1$ as all the bees will subsequently be recruited. If recruitment ceases in the first patch  (transient recruitment), $v_s = v_2(t_2) + \theta$ where $t_2$ is defined by $v_1(t_2) = \theta$. For the limiting cases discussed earlier, $v_s$ can be obtained analytically and numerically for the cases in between.

How best can the two patch hive discern between foraging events worth of a colony-wide recruitment event versus those for which the colony should stay in the hive? An effective foraging strategy would lead to the full hive being recruited when the foraging site is plentiful and transient recruitment when a discovered foraging site is of poor quality. Sometimes favorable sites may not lead to many initial recruits $v_1(0)$, and sometimes the  unfavorable state leads to many initial recruits $v_1(0)$ (Fig~\ref{fig5}b). Using signal detection theory, we show that if the colony is capable of tuning the recruitment threshold, they can facilitate foraging mostly at the favorable foraging sites and not at unfavorable sites.

Define the distribution of possible initial recruit fractions $v_1(0)$ in the favorable $f_{+}(v_1(0)) $ and unfavorable $f_{-}(v_1(0))$ cases with $v_1(0) \in [0,1]$. Given the initial distribution functions $f_{\pm}(v)$ and the long term limit of committed bees  $v_s \equiv g(v_1(0); \theta, \beta, D_w, D_v)$, we can compute the mean $v_s$ in both the favorable $\langle v_s \rangle_+$ and unfavorable $\langle v_s \rangle_- $ state using the formulas
\begin{align*}
\langle v \rangle_{\pm} | \theta, \beta, D_w, D_v = \int_0^1 g(v; \theta, \beta, D_w, D_v) f_{\pm}(v) d v.
\end{align*}

The long term foraging yield can be maximized across environments if many bees are recruited in the favorable environments and few bees are recruited in unfavorable environments. Define the long term yield as as the sum of the mean turn out of recruiters in the favorable state plus the mean of the fraction remaining uncommitted in the unfavorable state, representing the energy saved by not foraging an unfavorable site:
\begin{align}
J(\theta; \beta, D_v, D_w) = \int_0^1 (1-v_s) g_-(v_s) d v_s + \int_0^1 v_s g_+(v_s) d v_s = 1 - \langle v_s \rangle_- + \langle v_s \rangle_+.  \label{simpobj}
\end{align}
Note, we use Eq.~(\ref{simpobj}) for simplicity, but it is related (by a constant additive factor) to a more complicated objective function that provides bees foraging profitable sites ($+$) with twice the energy of those remaining in the hive and bees foraging unprofitable sites ($-$) with no energy:
\begin{align*}
\tilde{J}(\theta; \beta, D_v, D_w) &=2 \cdot \int_0^1 v_s g_+(v_s) d v_s +  \int_0^1 (1-v_s) g_-(v_s) d v_s + 0 \cdot \int_0^1 v_s g_+(v_s) d v_s +  \int_0^1 (1-v_s) g_-(v_s) d v_s  \\
& =  2 \cdot \langle v_s \rangle_+ + 1 - \langle v_s \rangle_+ + 0 \cdot  \langle v_s \rangle_- + 1 - \langle v_s \rangle_- = 2 - \langle v_s \rangle_-  +  \langle v_s \rangle_+ \\
&= J(\theta; \beta, D_v, D_w) + 1,
\end{align*}
and since the critical points are not affected by the constant shift, we retain Eq.~(\ref{simpobj}).

The threshold $\theta$ in $J(\theta; \beta, D_v, D_w)$ can be varied to maximize Eq.~(\ref{simpobj}), so the optimal threshold $\theta_c$ balances the risk of terminating recruitment to a high-yielding site due to a high threshold with recruitment to a perilous site through over-zealous recruiting (Fig~\ref{fig5}c). For most parameters,  $\theta_c \in [0.9, 0.16]$, tending to increase with $\beta$ (preventing rapid recruitment) and decreasing with $D_v$ (ensuring recruitment even when recruiters diffuse) (Fig~\ref{fig5}d).  The optimal threshold approaches a ceiling of $\theta = 0.16$ for higher values of $\beta$. This makes sense because when $\theta > 0.16$, most of the initial conditions (fraction of initial recrutiers) in favorable foraging environment $f_{+}(v)$ will lead to recruitment dynamics that fall below threshold too quickly. For lower values of $\beta$, the threshold $\theta$ should be decreased to prevent premature abandonment of the recruiting process in the case of high-quality foraging sites.

\subsection{Maximizing foraging yield}
\begin{figure}[t!]
\begin{center}    \includegraphics[width=15cm]{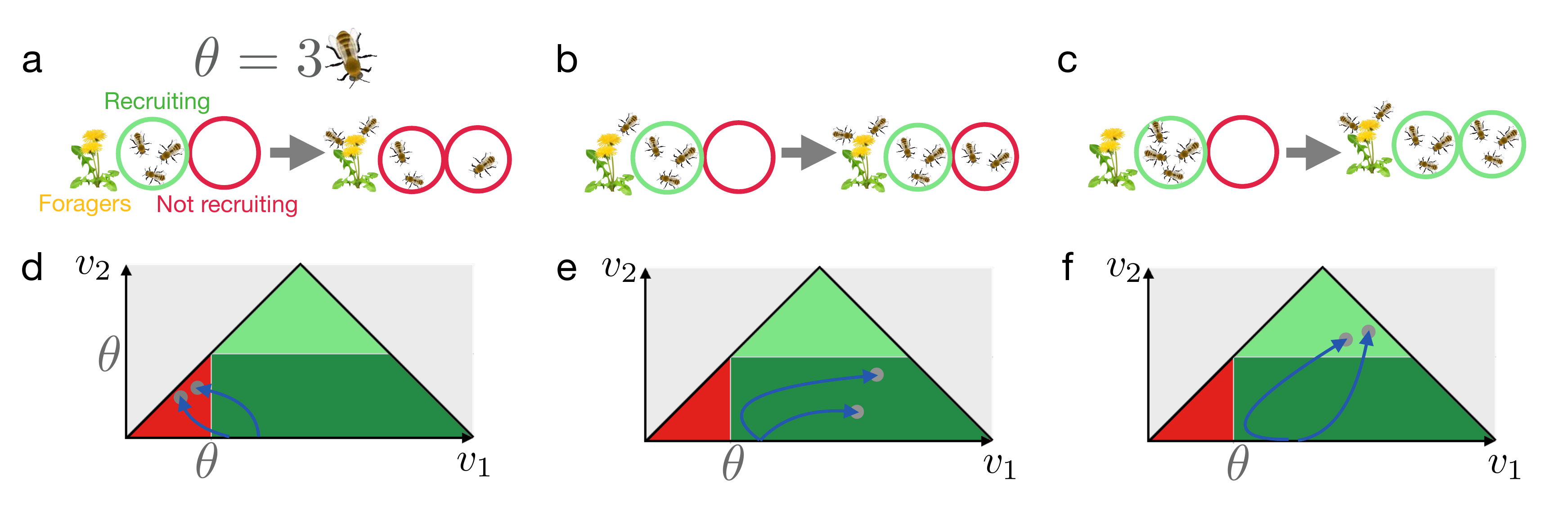}	\end{center} 	
\caption{{\bf Schematics of three different recruitment scenarios in a two patch hive model.} The recruitment threshold is three waggle dancing bees per patch. Green circle denotes an actively recruiting patch and red circle denotes a patch with no ongoing recruitment. {\bf a.} \bi{Transient recruitment:} Some waggle dancers diffuse to second patch. Both compartments are below recruitment threshold; {\bf b.}~\bi{Intermediate recruitment:} Recruitment begins in the first patch and some waggle dancers then diffuse to second patch but not enough to initiate recruitment in the second patch. Recruitment continues only in the first patch; {\bf c.}~\bi{Strong recruitment:} Recruitment begins in the first patch and enough waggle dancers diffuse to second patch to begin recruitment there. {\bf d,e,f.} Schematic of trajectories in phase plane corresponding to behaviors indicated in {\bf a,b,c}. If $v_1, v_2< \theta$ (red), recruitment ceases altogether, if $v_1>\theta>v_2$ (dark green), only the first patch is being recruited, and if $v_1, v_2>\theta$ (light green), both patches are being recruited. }
\label{fig6}
\end{figure}

Honey bee colonies not only benefit from consensus in foraging decisions, but must also implement strategies that result in sufficiently high foraging yields~\citep{dornhaus06}. We also examined how the diffusion of bees $(D_v, D_w)$ and the rate of exchange between the pure foraging and recruiting populations ($\nu, \mu$) shape the foraging yield of the colony.  For this section, we consider full model in Eq.~(\ref{discreteEqn}) where $N=2$. Under the initial conditions $v_1(0) > \theta, v_2(0) = 0$, Eq.~(\ref{discreteEqn}) reduces to  \begin{subequations}  \begin{align} 
\dot{w}_1 &= - \beta w_1 + D_w(w_2 - w_1); \hspace{13mm} \dot{w}_2 = D_w(w_1 - w_2);  \\ \dot{v}_1 &= - \mu v_1 + D_v(v_2 - v_1) + \nu u; \hspace{8mm} \dot{v}_2 = - \mu v_2 + D_v(v_1 - v_2); \\
\dot{u} &= \beta w_1 - \nu u + \mu (v_1 + v_2).
\end{align}  \label{eqn2p} \end{subequations} 
Based on dynamics of Eq.~(\ref{eqn2p}), there are three possible phases of colony behavior:

{\bf Transient recruitment:} In this case, the fraction of recruiters in first patch $v_1(t)$  decreases and falls below the threshold $\theta$ and recruitment ceases in both patches. This could occur due to a high rate of diffusion of waggling bees $D_v$ causing the waggling bees to quickly spread between patches and fall below the threshold (Fig~\ref{fig6}a),
or when most waggling bees become pure foragers (high $\mu$).  

{\bf Intermediate recruitment:} In this case, the first patch is fully recruited, but recruitment does not commence in second patch. This phase of recruitment occurs only when there is no movement of uncommitted bees $D_w = 0$.
This occurs when there is a low level of diffusion of wagglers ($D_v$) in the hive and the switch rate from waggling to foraging is much higher than the reverse. Too few wagglers move to the second patch and so the uncommitted bees in the second patch remain unrecruited (Fig~\ref{fig6}b). The dynamics eventually relaxes to the steady states of Eq.~(\ref{eqn2p}) given as
\begin{align}	\label{ssintrecruit}
 \bar{w}_1 =  0;  \ \	 
 \bar{w}_2 =  \frac{1-v_1(0)}{2};  \ \
\bar{u} =  \frac{\mu(1+v_1(0))}{2(\nu+\mu)}; \ \ 
\bar{v}_1 = \frac{\nu(1+v_1(0))(D_v + \mu)}{2(2 D_v + \mu)(\nu + \mu)}; \ \	 
\bar{v}_2 = \frac{\nu D_v (1+v_1(0))}{2(2 D_v + \mu)(\nu + \mu)};
\end{align}
with non-positive eigenvalues. This provides explicit bounds on model parameters that allow intermediate recruitment, since we need that $ \bar{v}_2 < \theta$ but $\bar{v}_1 > \theta$. 

{\bf Strong recruitment:} In this case, the fraction of recruiters increase in both patches and recruitment commences in second patch. The rate of diffusion of waggling bees ($D_v$) and the conversion of waggling bees to foraging bees ($\mu$) can be tuned to foster recruitment throughout the hive (Fig~\ref{fig6}c). Here, all the bees in the hive are eventually recruited to forage, maximizing the foraging yield of the colony when the foraging site is favorable. Note, the steady states of Eq.~(\ref{discreteEqn}) with $N=2$ in this case are
\begin{align}	\label{ss2patch}
 \bar{w}_1 =  \bar{w}_2 = 0; \ \
 \bar{u} = \frac{\mu}{\nu+\mu}; \ \
 \bar{v}_1 = \frac{\nu(D_v + \mu)}{(2 D_v + \mu)(\nu + \mu)}; \ \
 \bar{v}_2 = \frac{\nu D_v}{(2 D_v + \mu)(\nu + \mu)}.
\end{align} 
with non-positive eigenvalues. We note that in this case, the whole colony can be recruited even when the second patch is subthreshold. As noted previously, this occurs in the asymptotic limit of $D_v \to 0$. However, this kind of recruitment through just the first patch is slower than when both the patches are superthreshold, since it limits the portion of the uncommitted population that can be reached by recruiters.

\begin{figure}[t!]
\begin{center}    \includegraphics[width=15cm]{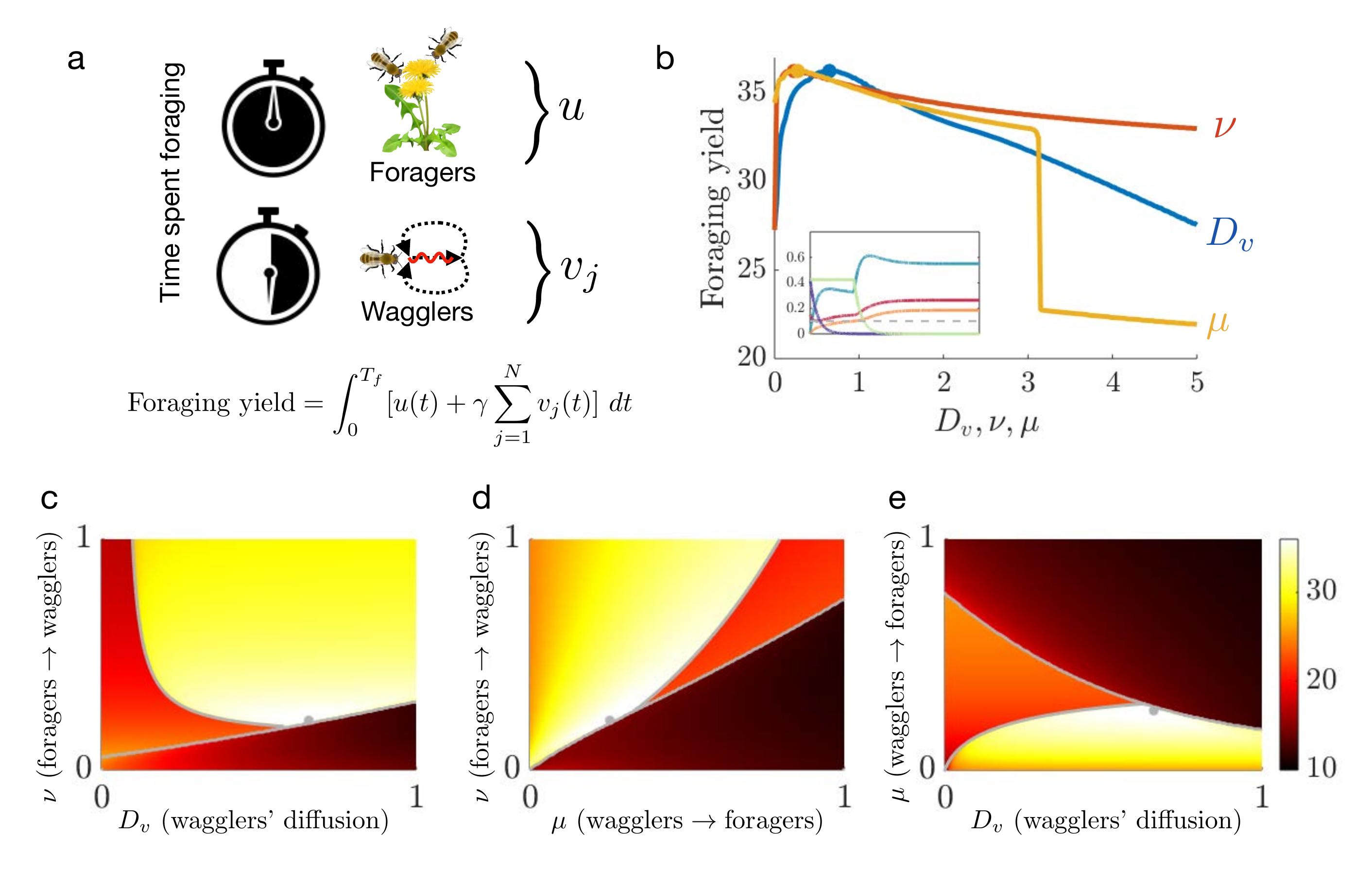}	\end{center} 	
\caption{  {\bf a.} Schematic showing calculation of foraging yield; {\bf b.} The foraging yield is maximized for a specific set of parameters $(D_v, \nu, \mu)$ when $\theta$, $\beta$, and $v_1(0)$ are fixed. For each of these curves, we fix the values of two other parameters such that they maximize the foraging yield at each point. Other parameters are $\beta = 1, D_w = 0, \theta = 0.1, v_1(0)=0.15$. The inset shows the recruitment dynamics at the optimal parameters, i.e $D_v = 0.66$, $\nu = 0.215$, $\mu = 0.255$.  {\bf c,d,e.} Heat-maps showing how the foraging yield varies with the model parameters. Abrupt changes in foraging yield occur when the colony switches between strong (yellow region), intermediate (red/orange region), and transient (dark region) recruitment regime. For each heat map, the parameter not on the axes is fixed such that the foraging yield with respect to the parameter is maximized. The maximum foraging yield on each map is represented with a grey dot.}
\label{fig7}
\end{figure}

Across these different phases of foraging and recruiting behavior, we calculate foraging yield using
\begin{align}	
J(\beta,D_v,D_w,\nu,\mu) = \int_{0}^{T_f} u(t) + \gamma \sum_{j=1}^N v_j(t) \ dt	. \label{objfun_forage}	
\end{align}
Bees in the foraging class $u$ spend most of their time foraging while the waggle dancing bees $v_j$ engage in foraging the fraction of the time between waggle runs (e.g., $\gamma = \frac{1}{2}$) (See Fig.~\ref{fig7}a for a schematic). The colony can tune the fraction of purely foraging bees and the fraction of recruiters by changing the rate they switch between these two behaviors ($\nu$ and $\mu$). The rate of diffusion of waggling bees $D_v$ can also be tuned to maximize the foraging yield (Fig.~\ref{fig7}b). 

The foraging yield of the colony closely depends on how colony members are recruited. Strong recruitment results in the highest foraging yield as there are greater number of bees foraging at least half of the forage cycle. Transient recruitment results in the lowest foraging yield. The heat maps for the foraging yield clearly delineates the regions corresponding to different recruitment phases (Fig~\ref{fig7}.c,d,e). Clearly there is an optimal rate of diffusion for wagglers $D_v$ that maximizes the foraging yield, but the dependence upon the role switching rates $(\mu, \nu)$ suggests both should be relatively low, so recruits initially dwell in the foraging class momentarily before transitioning to waggling. Note the optimal parameter set is very close to the boundaries between all three regions of behavior. There is a fine balance between devoting sufficient wagglers to recruitment and maximizing the fraction of active foragers (tuning $\nu$ and $\mu$) diffusing fast enough and spreading the population too thin across patches (tuning $D_v$).

We can partition parameter space into different types of recruitment by analyzing the solution of Eq.~(\ref{eqn2p}) in the limit of slow uncommitted bees ($D_w \to 0$). Here, we can can solve Eq.~(\ref{eqn2p}) explicitly, so
\begin{align*}
v_1(t) =&\frac{D_v  \left(v_1(0) \left(-2 \mu  (\beta + \nu )+\beta  \lambda +8 D_v^2-4 D_v (\beta + \nu -2 \mu )+2 \mu^2\right)+\beta  \nu \right)}{2 (2 D_v - \nu ) (2 D_v+\mu ) (-\beta +2 D_v+\mu )}e^{-t (2 D_v+\mu )} \\
   & + \frac{(D_v - \nu ) \left(v_1(0) \beta \nu+2 v_1(0) \beta  \mu -2 v_1(0) \nu  \mu -2 v_1(0) \mu ^2-\beta \nu \right)}{2 (2 D_v -\nu )(\nu +\mu ) (\beta - \nu -\mu )}e^{ -(\nu + \mu )t}   \\
   & - \frac{(v_1(0) - 1) \nu (\beta - D_v - \mu )}{2 (\beta -\nu -\mu ) (\beta - 2 D_v - \mu )}e^{- \beta t}  +\frac{(v_1(0)+1)\nu  (D_v +\mu )}{2 (2 D_v+\mu )(\nu +\mu )},
\end{align*}
and transient recruitment occurs where $v_1(t) < \theta $, allowing us to separate parameter space based on this condition.

To denote strong recruitment, we can analyze the steady states of Eq.~(\ref{ssintrecruit}) in the limit of small diffusion of uncommitted bees ($D_w \to 0$),
\begin{align*}
\bar{v}_2 = \frac{\nu D_v (1+v_1(0))}{2(2 D_v + \mu)(\nu + \mu)}.	
\end{align*}
Recruitment commences in the second patch when $v_2(t) \geq \theta$, thus setting $ \frac{\nu D_v (1+v_1(0))}{2(2 D_v + \mu)(\nu + \mu)} > \theta$ provides a condition for strong recruitment to exist in a hive with two compartments (Fig.~\ref{fig7}c,d,e).  \\

\noindent
{\bf Single patch hive.} To better understand the dynamics of waggler and forager interactions, we also examined the case of a hive with single patch for comparison. For $N=1$, under the initial condition $v_1(0) > \theta$ and for as long as $v_1(t) > \theta$, Eq.~(\ref{discreteEqn}) reduces to
\begin{align}
w' = - \beta w,		\ \	v'= - \mu v+ \nu u,  \ \ u' =  \mu v- \nu u + \beta w, 	\label{eqn1patch}
\end{align}  
which can be solved to obtain
\begin{subequations} 	\begin{align}
w(t)  =& e^{-\beta t} \left(1-v(0) \right) \\ 
v(t) =& \frac{\nu}{\nu+ \mu} + \frac{\nu (1-v(0))}{\beta - \nu- \mu}  e^{-\beta t} - \frac{v(0) \mu (\nu+ \mu) + \beta(\nu - v(0)\nu - v(0) \mu )}{(\beta - \nu - \mu)(\nu + \mu)}  e^{-(\nu + \mu )t}  \\
u(t) =& \frac{\mu}{\nu+ \mu} + \frac{(1-v(0)) (\mu - \beta)}{\beta - \nu - \mu}  e^{-\beta t} + \frac{v(0) \mu (\nu + \mu) + \beta(\nu - v(0)\nu - v(0) \mu )}{(\beta - \nu - \mu)(\nu + \mu)}  e^{-(\nu + \mu )t}.
\end{align}  \label{singlesoln}	\end{subequations}
We calculate foraging yield as 
\begin{align} 
J(\beta , \nu, \mu,T) =& \frac{v_1(0) \mu(\nu +\mu )+\beta  (\nu -v_1(0)\nu -v_1(0) \mu )}{(\nu +\mu )^2 (-\beta +\nu +\mu )}(\gamma -1)(1-e^{-T (\nu +\mu )})  \\
&  + \frac{(v_1(0)-1) (\beta -\gamma \nu -\mu )}{\beta(\beta -\nu -\mu)}(1-e^{-\beta T}) +\frac{T(\gamma \nu+\mu) }{\nu+\mu }. \label{forageJone}
\end{align}
Clearly we see that the problem of optimizing foraging yield involves balancing multiple terms that arise from the exchange of bees between each of the three populations. The final term represents the long term yield from the combination of wagglers ($\gamma \nu$) and foragers ($\mu$). The first two terms represent the contributions to forager from the initial transient dynamics while bees are reaching their equilibrium fractions. Clearly, if $T$ is sufficiently large, it is best to make $\mu$ as large as possible, while still allowing continued recruitment until the whole colony is recruited. Thus, the limit on increasing foragers is primarily provided by the strong need for recruiters in the initial transient stage.

Inherent in Eq.~(\ref{eqn1patch}) is the assumption that $v(t) > \theta$ $\forall t>0$. Thus, we add a constraint that the minimum of $v(t) $ is greater than $\theta$,
\begin{align}	\label{singlemin}
v(t^*) = \frac{\lambda }{\lambda +\mu }-\frac{(1- v(0)) \lambda  \left(\frac{(1-v(0)) \beta  \lambda }{\beta  (\lambda  -v(0) \lambda -v(0) \mu )+v(0) \mu  (\lambda +\mu )}\right)^{\frac{\beta}{-\beta +\lambda +\mu }}}{\lambda +\mu } > \theta.
\end{align}
Thus, foraging yield is maximized by optimizing Eq.~(\ref{forageJone}), subject to the constraint Eq.~(\ref{singlemin}).

In a patchy hive model, colonies that adapt individual movement to the properties of hive are more effective in carrying information throughout the hive quickly. We find two different strategies for effective propagation of social information in hive based on the number of hive patches. In hives with fewer patches, rapidly moving wagglers can disperse information throughout the hive. On the other hand, the colony relies more on the movement of uncommitted bees for rapid information transfer in larger hives. Movement of bees also strongly shapes foraging yield, which on short timescales is maximized when the colony is rapidly recruited, foregoing the devotion of bees to the foraging class. On longer timescales, a large class of wagglers is less important, as long as it is sufficient to complete a total recruitment of the hive. Our analysis demonstrates that the spatial extent of the hive is a strong driver of the optimal strategy both for rapid recruitment as well as the allocation of bees to different roles.

\section{Continuum model of position in the hive}

\begin{figure}[t!]
\begin{center}    \includegraphics[width= 15 cm]{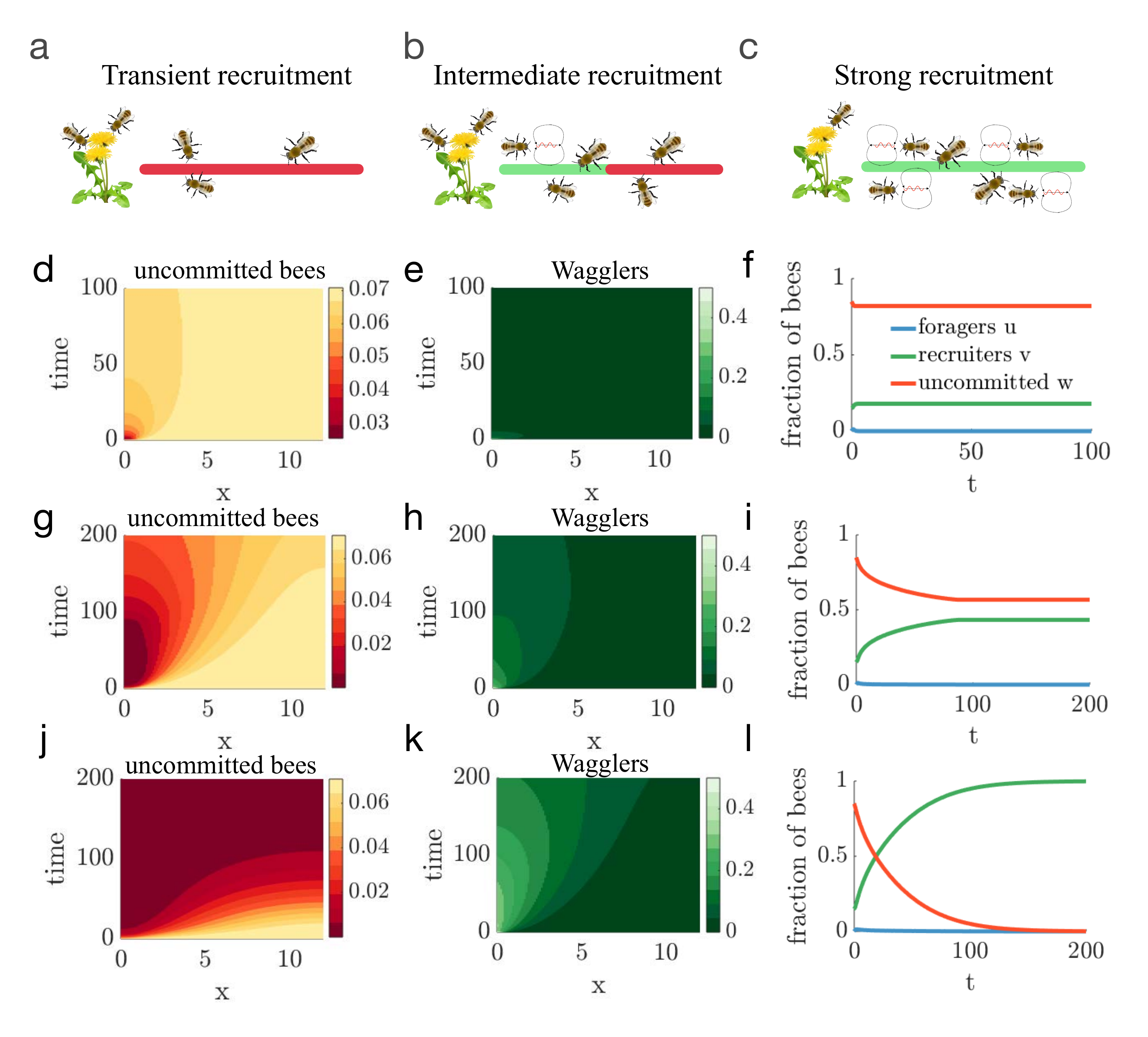}	\end{center} 	
\caption{  {\bf a,b,c.} Schematic of three different recruitment scenarios in a line hive model representing committed (green) and uncommitted (red) bees.  {\bf d,e.}~Changes in number of uncommitted and waggle dancing bees in the hive in a transient recruitment scenario. {\bf f.} Change with time in number of bees in each class in a transient recruitment scenario $\beta = 1, D_w = 0.1, \theta =0.1, \nu = 2, \mu = 0, D_v = 1$. {\bf g,h.}~Changes in the number of uncommitted and waggle dancing bees in the hive in an intermediate recruitment scenario.  {\bf i.}~Change with time in number of bees in different class in an intermediate recruitment scenario $\beta = 1, D_w = 0.1, \theta = 0.1, \nu = 2, \mu = 0, D_v = 0.1$. {\bf j,k.}~Change with time in number of bees in different class in a strong recruitment scenario. {\bf l.}~Change with time in number of bees in different class in a strong recruitment scenario $\beta = 1, D_w = 1, \theta =0.1, \nu = 2, \mu = 0, D_v = 0.1$.}
\label{fig8}
\end{figure}

We have shown that increasing the size of a hive can have strong consequences on the ability of a colony to complete a total recruitment. Indeed, as the hive grows, the motion of uncommitted bees becomes more important. Now we consider the continuum limit of a spatially distributed hive, and analyze how efficient foraging strategies are then shaped by the size of the hive and the parameters of movement and recruitment. Starting with a patch model, take the right edge of each patch to be $x_j = j \cdot \Delta x$ where $\Delta x = L/N$, the length of the hive $L$ divided by the number of patches $N$. Fixing $L$ and taking the limit $N \to \infty$, then $x \in [0, L]$. If we also divide the diffusion coefficients $D_c$ ($c = w,v$) by $(\Delta x)^2$ so $\hat{D}_c = D_c/(\Delta x)^2$, then $\hat{D}_c (c_{n+1} - 2 c_n + c_{n-1}) \to  D_c c_{xx}(x,t)$ on the interior. At the boundaries, we take a no flux condition, so $w_x(0,t) = w_x(L,t) = 0$. Thus, the population model for non-committed and waggling bees in the hive evolves on a one dimensional line segment. The spatiotemporal evolution of the waggling $v(x,t)$, uncommitted $w(x,t)$ and foraging $u(t)$ bees evolves according to a piecewise smooth coupled system of partial differential equations (PDE: schematic in Fig.~\ref{fig1}d):
\begin{subequations} 	\begin{align}
\frac{\partial w}{\partial t}  =& D_w  w_{xx}  - \beta  H[v(x,t) - \theta] w(x,t),  \qquad w_x(0,t) = w_x(L,t) = 0, \\
\frac{\partial v}{\partial t}  =& D_v  v_{xx} - \mu v(x,t),     \hspace{12mm} -D_v v_x(0,t) =  \nu u(t), \qquad v_x(L,t) = 0, \\
\frac{du}{dt}  =& \beta \int_{0}^{L} H[v(x,t) - \theta] w(x,t) \ dx + \mu \int_{0}^{L} v(x,t) \ dx  - \nu u(t)  
\end{align}  \label{pdeeqn} 	\end{subequations}
Waggling recruiters enter the hive at one of the ends $(L=0)$ according to the partially boundary condition (arising from the continuum limit of the entry point being solely in the first patch) and recruit uncommitted hive mates as long as their local density exceeds a threshold $\theta$. To measure the efficacy of recruitment, we will compute the two same measures as before as population motion, foraging, and recruitment parameters are varied: time to recruit $90 \%$ of the hive population and foraging yield.  Analogous to the discrete patch hive, colonies may tune their diffusion rates ($D_v$ and $D_w$) to alter the rate of information propagation about foraging site availability throughout the hive.

Note, the form of the population density boundary conditions for the PDE is essential for conservation of bees. The total number of bees $b(t) = \int_0^L w(x,t) dx + \int_0^L v(x,t) dx + u(t)$ is constant:
\begin{align*}
\frac{db}{dt} = & D_w \int_0^L w_{xx}(x,t) dx - \beta \int_0^L H[v(x,t) - \theta] w(x,t) dx + D_v \int_0^L v_{xx}(x,t) dx - \mu \int_0^L v(x,t) dx \\
& + \beta \int_{0}^{L} H[v(x,t) - \theta] w(x,t) \ dx + \mu \int_{0}^{L} v(x,t) \ dx  - \nu u(t) \\
= & D_w (w_x(L,t) - w_x(0,t)) + D_v(v_x(L,t) - v_x(0,t)) - \nu u(t) = \nu u(t) - \nu u(t) = 0.
\end{align*}

As we do not include spontaneous commitment and abandonment, we consider the role of an external foraging source as providing an initial fraction of recruiters at the entrance of the hive $(L=0)$ at time $t=0$ such that $\int_0^L v(x,0) dx = \int_0^L \zeta \delta (x-0^+) dx  = \zeta$. The remaining uncommitted bees are initially spread evenly in the hive $w(x,0) = \frac{1-\zeta}{L}$ where $L$ is the length of the hive, and are recruited in the regions where waggling population is above recruitment threshold $\theta$, initially just near the hive entrance. 

When uncommitted bees are recruited, the fraction of wagglers in the hive can increase. However, if the switching rate from pure foragers to recruiters is low, quick diffusion of wagglers can lead to the waggler population falling subthreshold throughout the hive and recruitment ceases. The three different types of recruitment (transient, intermediate and strong) persist in the continuous model, though the distinction between transient and intermediate is less pronounced. Fig.~\ref{fig8} shows how the hive population and densities evolve in each of these three different recruitment scenarios. When wagglers diffuse rapidly, their initial population quickly spreads throughout the hive such that the waggler density becomes subthreshold everywhere in the hive (Fig.~\ref{fig8}d,e,f). As in the discrete patch model, excessive movement of recruiters limits their impact. 
When the diffusion of wagglers is slower and the recruitment rate is low, recruitment is sustained for a brief period of time before the waggler density becomes subthreshold resulting in fewer than half of the colony being recruited (Fig.~\ref{fig8}g,h,i).
On the other hand, if uncommitted bees move rapidly while wagglers diffuse slowly, the waggler density remains superthreshold in a region of the hive near the entrance, eventually recruiting the entire hive (Fig.~\ref{fig8}j,k,l).  This is similar to the discrete hives with a large number of patches, where only way to recruit the majority of the hive population is through considerably slow diffusion of recruiters $D_v \to 0$ and to allow the uncommitted bees to meet this population. This illustrates how a critical threshold on the fraction of recruiters needed to recruit foragers can lead to a requirement of the localization of recruitment on dance floors to achieve widespread recruitment of the uncommitted population, especially in large hives where maintaining a population of superthreshold recruiters throughout the hive is not possible.

As before, efficient recruitment requires balancing a rapid propagation of the recruitment signal without the population of recruiters becoming too diffuse. To study the problem more closely, we start by considering some limiting cases, as in the discrete patch hive model. In the first of these cases, we consider a model with no pure foraging class. \\
\vspace{-3mm}

\noindent {\bf No foraging class.} Taking the limits $\nu \to \infty$ and $\mu \to 0$, any committed bee instantaneously becomes a recruiter and remains that way indefinitely. Thus, bees are either recruiting or uncommitted:
\begin{subequations}  \label{contnoforag}	\begin{align}
\frac{\partial w}{\partial t}  =& D_w  w_{xx}  - \beta  H[v(x,t) - \theta] w(x,t),   \qquad \qquad w_x(0,t) = w_x(L,t) = 0 \\
\frac{\partial v}{\partial t}  =& D_v  v_{xx},  \qquad \qquad D_v v_x(L,t) =  D_v v_x(0,t) +  \beta \int_{0}^{L} H[v(x,t) - \theta] w(x,t) \ dx = 0
\end{align} 	\end{subequations}
Movement of committed and uncommitted individuals play a crucial role in determining whether the hive population is recruited entirely and how quickly.

\begin{figure}[t!]
\begin{center}    \includegraphics[width=15cm]{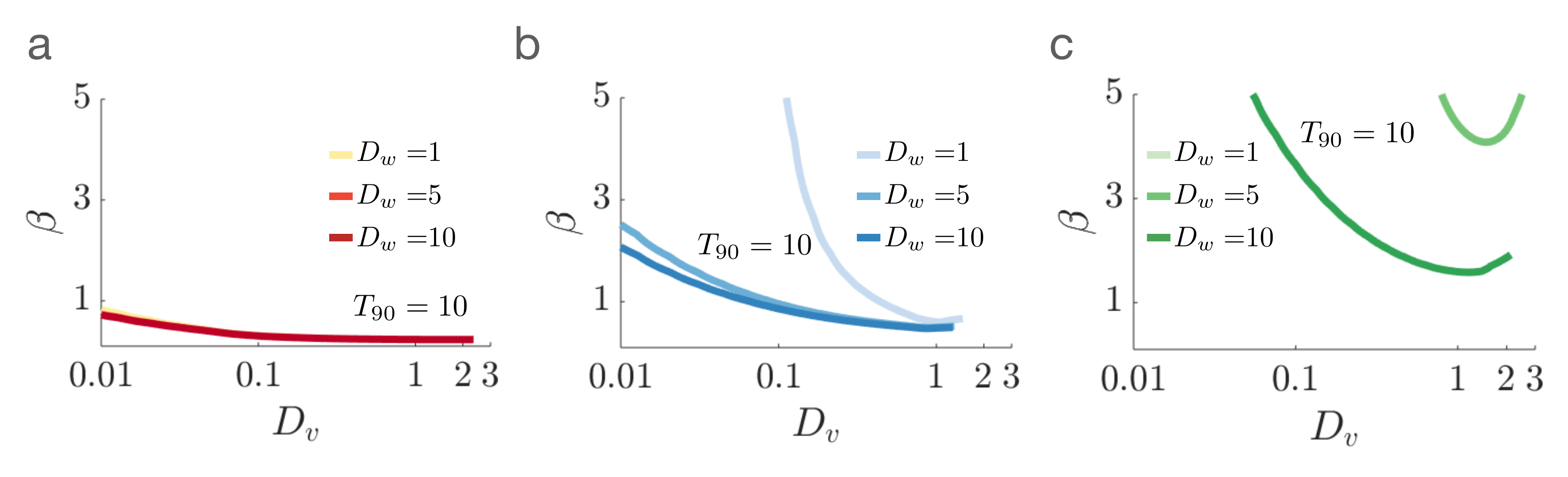}	\end{center} 	
\caption{Level curves for time to recruit $90 \%$ of the hive ($T_{90} = 10$) as they vary with the parameters of the continuous model without foragers, Eq.~(\ref{contnoforag}). Note for parameter values above (below) the curve, the time to recruit $90\%$ is less than (more than) $T_{90} = 10$. We show the curve $T_{90} = 10$ varies with rate of recruitment $\beta$, diffusion coefficient for wagglers $D_v$, and diffusion of uncommitted bees $D_w$ for three different hive lengths: {\bf a.} $L =2$ {\bf b.} $L=5$ {\bf c.} $L=10$. }
\label{fig9}
\end{figure}

We examine specifically how the two diffusion rates $(D_v, D_w)$ shape the time to recruit the colony (Fig.~\ref{fig9}). In contrast to the finite discrete patch case, here the hive length $L$ is the free parameter determining the relative extent of the hive. When the hive is short, the movement of uncommitted bees impacts the dynamics of recruitment weakly. This is because any movement of the uncommitted bees tends to quickly spread their population distribution across the hive. The time to recruit $90 \%$ of the hive thus does not noticeably vary with changes in $D_w$ for $L = 2$ (Fig.~\ref{fig9}a). As the length of hive increases, diffusion of uncommitted bees significantly reduces the time to recruit nearly all the colony (Fig.~\ref{fig9}b,c). Interestingly, for the longest hive length $L=10$ we consider, the level curve is non-monotonic in $D_v$, since excessive diffusion can spread the recruiter population thin. Analogous to the patch hive model, we see here that for a fixed value of $\beta$, there is an optimal balance of recruiter and uncommitted bee diffusion that minimizes the time to near total recruitment. Also, level curves for $T_{90}$ do not exist for higher values of diffusion rates $D_v$, since recruitment terminates early due to the rapid diffusion  of the waggler population causing them to locally fall below threshold. For longer hive lengths $L=10$, level curves for $T_{90}$ do not exist for low $D_w$. \\
\vspace{-3mm}
 
\begin{figure}[t!]
\begin{center}    \includegraphics[width=15cm]{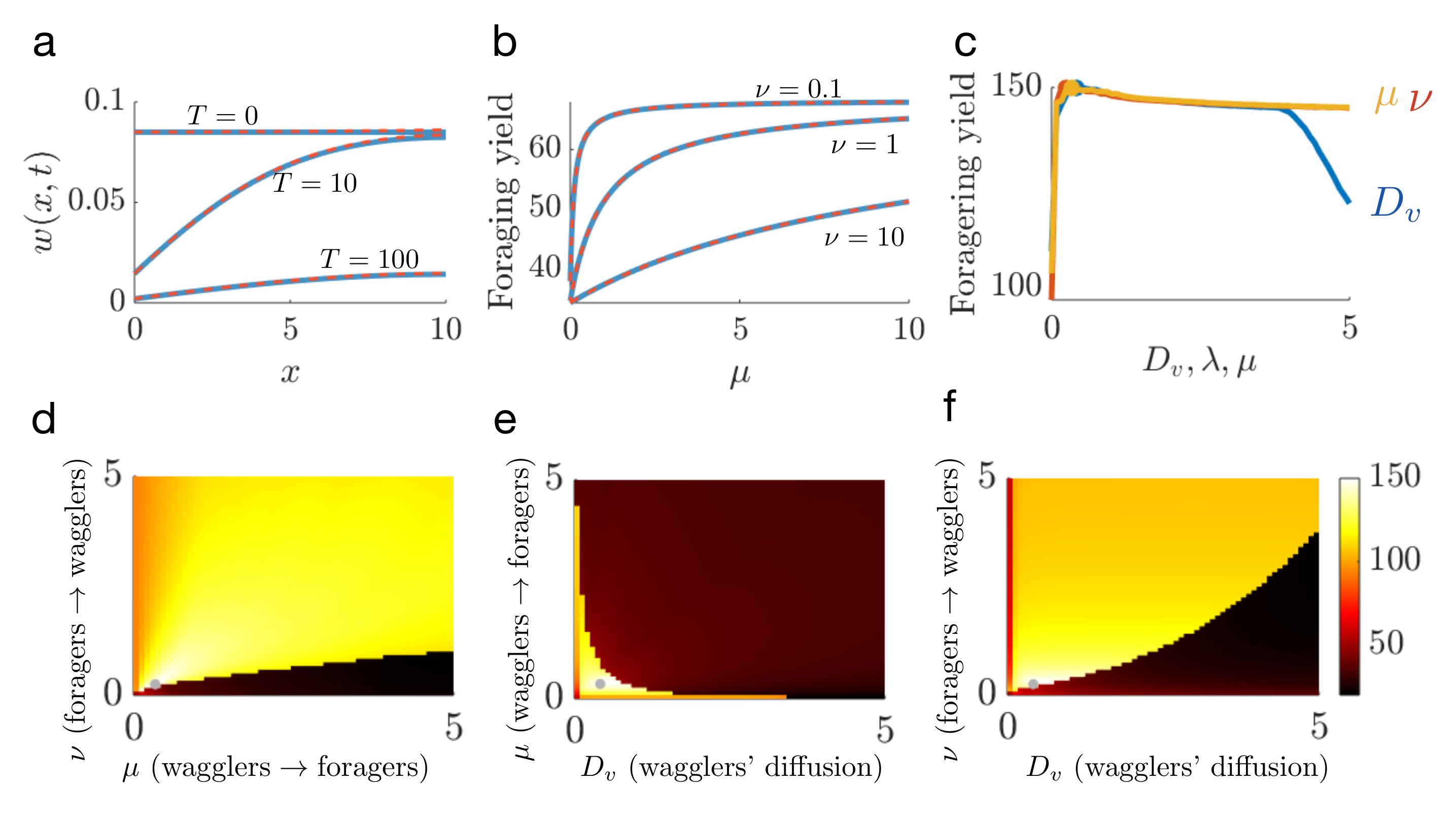}	\end{center} 	
\caption{{ \bf a.} Distribution of wagglers in the hive obtained by solving Eq.~\ref{Dv0pde} at times: $T = 0,10,100$. {\bf b.} Foraging yield as it varies with increase in foragers $(\mu)$ and wagglers $(\nu)$ when $\beta =1, D_w =1, D_v =0$. Blue solid lines are the Fourier series solutions and dashed red lines are numerically computed solutions. {\bf c.} Foraging yield varies with the model parameters in the continuum hive model. There are parameter optima that maximize the foraging yield. For each of the curve, we fix the values of two other parameters such that they maximize the foraging yield ($\beta =1$, $D_w = 0.01$, $\theta = 0.1$, $L = 2$).  {\bf d,e,f.} Heat maps show how the foraging yield delineate the regions in parameter space corresponding to strong and transient recruitment. For each heat map, the parameter not on the axis is fixed such that the foraging yield with respect to the parameter is maximized ($D_v =  0.4167 $, $\nu = 0.25$, $\mu = 0.3390$). The grey dot shows the maximum yield.}
\label{fig10}
\end{figure}

\noindent{\bf Slow waggler movement.} We now consider the role of pure foragers in shaping the foraging yield of the colony. To better understand the exchange of roles throughout the colony, and how this shapes recruitment, we consider the limit of slow moving recruiters $D_v \to 0$, so the recruitment term becomes a boundary condition in our model, and the left boundary condition of the uncommitted population becomes partially absorbing (Robin boundary condition). The other two equations are piecewise linear differential equations: 
\begin{subequations} 	\begin{align}
\frac{\partial w}{\partial t}  =& D_w  w_{xx},   \qquad \qquad D_w w_x(0,t) -  \beta H[v^0(t) - \theta] w(0,t) = w_x(L,t) = 0 \\
\frac{\partial v^{0}}{\partial t}  =& \nu u(t) -\mu v^{0}(t),	\\
\frac{\partial u}{\partial t}  =& \mu v^{0}{t} - \nu u(t) +  \beta H[v^0(t) - \theta] w(0,t).
\end{align}  \label{Dv0pde}	\end{subequations}
Note, the partially absorbing boundary condition associated with $w$ emerges most clearly when taking the $D_v \to 0$ prior to taking the continuum spatial limit. Assuming $v^0(t)> \theta$ indefinitely,  we can solve this system of equations to obtain a Fourier series solution for the density of uncommitted bees, 
\[ w(x,t) = \sum_{n=0}^{\infty} a_n e^{-D_w \lambda_n t} \left( \frac{ D_w \sqrt{\lambda_n}}{\beta} \cos(\sqrt{\lambda_n} x) +\sin(\sqrt{\lambda_n} x)  \right) \]
where the eigenvalues $\lambda_n$ are defined implicitly such that $\tan(\sqrt{\lambda} L) = \frac{\beta}{D_w \sqrt{\lambda}} $ and 
\[ a_n  = \frac{ \int_{0}^L \frac{1-\zeta}{L} \left( \frac{ D_w \sqrt{\lambda_n}}{\beta} \cos(\sqrt{\lambda_n} x) +\sin(\sqrt{\lambda_n} x) \right) \ dx}{\int_{0}^L \left( \frac{ D_w \sqrt{\lambda_n}}{\beta} \cos(\sqrt{\lambda_n} x) +\sin(\sqrt{\lambda_n} x) \right)^2 \ dx} 
= \frac{1-\zeta}{L} \left( \frac{2 \beta^2}{ \sqrt{\lambda_n} (L \beta^2 + L D_w^2 \lambda_n + \beta D_w)}\right), \]
which matches simulations of the full system very well (Fig.~\ref{fig10}A). The waggler density at the left boundary is 
$w(0,t) = \frac{D_w }{\beta} \sum_{n=0}^\infty a_n \sqrt{\lambda_n} e^{-D_w \lambda_n t}$, and so Eq.~(\ref{Dv0pde}b,c) becomes
\begin{align*}
 v^{0}_{t} = \nu u(t) -\mu v^{0}(t)	;	\ \ 
u_{t}  = \mu v^{0}(t) - \nu u(t) +  D_w \sum_{n=0}^\infty a_n \sqrt{\lambda_n} e^{-D_w \lambda_n t}.
\end{align*}
Solving this system of differential equations gives the series solution,
\begin{align*}
v^{0}(t) =&  \frac{\nu \zeta }{\mu +\nu}  +   \frac{\nu }{\mu +\nu}\sum_{n=0}^\infty \frac{a_n}{\sqrt{\lambda_n}} +\frac{e^{-(\mu + \nu)t}  }{\mu +\nu} \left(  \zeta \mu - D_w \nu  \sum_{n=0}^\infty \frac{a_n \sqrt{\lambda_n}}{D_w \lambda_n - \mu -\nu} \right)  + \sum_{n=0}^\infty \left( \frac{a_n \nu e^{- \lambda_n D_w t} }{\sqrt{\lambda_n}(D_w \lambda_n - \mu -\nu)}  \right)  \\
u(t) =&  \frac{\mu  \zeta }{\mu +\nu} +  \frac{\mu }{\mu +\nu} \sum_{n=0}^\infty \frac{a_n}{\sqrt{\lambda_n}}  - \frac{e^{-(\mu + \nu)t} }{\mu +\nu} \left(  \zeta \mu - D_w \nu  \sum_{n=0}^\infty \frac{a_n \sqrt{\lambda_n}}{D_w \lambda_n - \mu -\nu} \right) + \sum_{n=0}^\infty \left( \frac{a_n e^{- \lambda_n D_w t} (\mu - D_w \lambda_n)}{\sqrt{\lambda_n}(D_w \lambda_n - \mu -\nu)}  \right).
\end{align*}
As before, we are interested in how the diffusion of bees ($D_v, D_w$) and rate of switching between pure foraging and recruiting influence foraging yield of a colony in the continuum model. Foraging yield is calculated using a continuous version of Eq.~\ref{objfun_forage},
\begin{align}
J(\beta,D_w,\nu,\mu) = \int_{0}^{T_f} \left( u(t) + \gamma \int_{0}^{L} v(x,t) \ dx \right) \ dt. \label{contfun_forage}
\end{align}
Thus, in this limiting case, we approximate the foraging yield using a finite number of terms from the Fourier series:
\begin{align*}
J(\beta,D_v,D_w,\nu,\mu,T)  =& \frac{ \zeta T (2 \mu +\nu )}{2 (\mu +\nu )} + \frac{1}{2} \sum _{n=0}^N \frac{\nu  a_n (1 -e^{-T D_w \lambda _n})}{D_w \lambda _n^{3/2}  \left(D_w \lambda _n-\mu -\nu \right)} +\sum _{n=0}^N \frac{a_n (1-e^{-T D_w \lambda _n}) \left(\mu -D_w \lambda_n\right)}{ D_w \lambda _n^{3/2} \left(D_w \lambda _n-\mu -\nu \right)} + \\
&\frac{T (2 \mu +\nu) \left(\sum _{n=0}^N \frac{a_n}{\sqrt{\lambda _n}}\right)}{2 (\mu +\nu )} + \frac{ (1- e^{-T (\mu +\nu )}) \left(\nu  D_w \sum _{n=0}^N \frac{a_n \sqrt{\lambda _n}}{D_w \lambda _n-\mu -\nu }-  \zeta \mu \right)}{2 (\mu +\nu )^2},
\end{align*}
and, for instance, we see that as the rate $\mu$ of transition to foraging is increased, so does the foraging yield (Fig.~\ref{fig10}B). Increasing the rate at which the recruiter population is increased decreases the foraging yield in this case, since it does not increase the rate of recruitment and limits the fraction of recruited bees devoted to foraging.  \\

\noindent
{\bf General parameters.} Returning to the general model, Eq.~(\ref{pdeeqn}), we now study more closely how foraging success depends on the movement of bees and their propensity to transition between the foraging and waggler class. As in the case of patchy hives, bees in the foraging class spend all of their time foraging while the wagglers divide their time between foraging and waggle dancing in the hive. As in discrete patches, foraging yield can be maximized by balancing the amount of time spent recruiting versus foraging and tuning the movement of committed and uncommitted bees accordingly to prevent spreading recruiters too thin while propagating information about food availability as quickly as possible. If all committed bees were to become pure foragers, the foraging yield gained from those particular bees would increase but recruitment in the hive would cease as there would be no wagglers left to continue recruiting in the hive. We see this scenario in the heatmap of foraging yield in Fig.~\ref{fig10}d,e, where high switch rate of bees from recruiting to foraging (high $\mu$) limit the gains of the colony in foraging. Clearly, it is best to limit the diffusive movement of the recruiters as well as the rates of transition between recruiter and forager. The apparent advantage seems to be that some recruited bees temporarily forage before becoming recruiters. Noticeably also, if $D_v$ is made too high or $\nu$ (the transition rate to becoming a recruiter) too low, then there is a significant drop in foraging yield indicative of recruitment terminatation (Fig.~\ref{fig10}d and e; as in Fig.~\ref{fig7}c,d).

Increasing the number of recruiters too much (large $\nu$ or small $\mu$) can also reduce foraging yields as the colony underutilizes its ability to have bees that purely forage. Once a significant proportion of the colony is recruited, there is less need for the waggling population.

Overall, the spatially-extended model of the hive leads to recruitment dynamics and qualitative results that are similar to the patch-based model of a hive. In general, increasing the spatial extent of the hive (either longer or with more patches), leads to the slower recruitment of the non-committed bees, so motility of the uncommitted bees is crucial for the eventual recruitment of the whole colony. Moreover, a balance of pure foragers and recruiters is necessary to maximize the foraging yield of the whole colony. Crucially, enough recruiters must be cycled into the population to maintain a steady stream of recruited bees, so that the recruiter population does not fall below threshold.

\section{Discussion}
The spatial geometry of a nest's interior can considerably impact the motility of social insects within and ultimately how information about the inside and outside world are propagated throughout~\citep{pinter15}. As honeybees communicate information about foraging sites and recruit nest-mates via waggle dancing, the spatial properties of their nest can affect colony-wide recruitment~\citep{tautz96}. We illustrate the effects of nest site geometry on the rate of recruitment and yield of a colony of foraging honeybees. Inspired by the importance of curbing positive feedback in the recruitment process~\cite{pagliara18} and nonlinearity in collective communication~\citep{passino08}, we used a threshold nonlinear function of wagglers concentration to model the rate of recruitment of uncommitted bees.

Our spatially-extended model of collective foraging decisions in honeybee colonies shows the efficacy of recruitment depends on the ability of the colony to tune the movement of waggle dancing and uncommitted individuals according to the size or length of the hive, ensuring the recruiting population is neither spread too thin nor moving too slowly as to unnecessarily limit recruitment rate. When the hive consists of only two patches or when the length of the hive is short in the continuous case, diffusion of uncommitted bees is superfluous and the colony relies mostly on the movement of wagglers to recruit everyone. However for larger hives, it is not feasible to maintain a population of recruiters above threshold throughout the hive and the diffusion of uncommitted bees is important for their continued recruitment. We also found that for larger hives, the only way colony can recruit the hive is by localization of recruitment to the region near the entrance through very slow movement of waggle dancing bees supporting the theory of localized dance floors in hives observed in the field~\citep{tautz97,seeley92}.

We also found that the threshold nonlinear recruiting function generates a model that is robust against recruitment to spurious foraging sites. Using the principles of signal detection theory, bees can avoid perilous and low yielding foraging sites while encouraging full colony recruitment to an ample foraging site. Such signal detection principles are known to be utilized by individual bees close to flowers, trying to determine their yield based on multiple flower attributes~\citep{leonard11}. Our results are reminiscent of the types of signal detection mechanism common to the nervous system, whereby only sufficiently strong stimuli lead to a population level response of sensory areas of the brain~\citep{verghese01}. Recently, there has been a growing interest in bringing traditional notions of signal detection theory from neuroscience and psychology into the arena of ecological phenomena such as foraging and predator-prey interactions~\citep{leavell19}. Not only might an animal use signal detection mechanisms for decision-making based on observations of their environment, they may also use these principles to help determine how to use information communicated to them by conspecifics. In addition to providing a novel perspective on collective foraging decisions, our model is also convenient in its piecewise linear formulation whereby it can be solved explicitly in a number of contexts to shed light on the qualitative boundaries between recruitment and termination.

Through the resulting time series of recruited bees, spatial properties of the hive also shape the foraging yield. Timely dissemination of information and hence recruitment to a foraging site is crucial in optimizing the foraging yield of a colony~\citep{seeley88}. Slow recruitment delays the harvest of resources, leading to lost time during which food sources may deteriorate or be consumed by competitors~\citep{seeley95,edge12,raihan14}. Furthermore, the colony should properly balance time spent foraging and recruiting uncommitted bees through a division of labor process~\citep{seeley83,dreller98}. We have demonstrated here that a split and substantial proportion of pure foragers and recruiters, who spend some time foraging, leads to the maximal foraging yield of a colony. Devoting all or most committed individuals to purely foraging leads to a situation where too few wagglers are available to recruit the remaining uncommitted individuals from the hive. In this case, while the recruited bees are making efficient use of their time, they are underutilizing the foraging potential of the remaining uncommitted bees. On the other hand, it is unnecessary to have every committed bee split their time between recruiting and foraging, since at later times no recruiting is needed at all. In future instantiations of our model, we could consider recruiters who are responsive to the need for waggling, based on the fraction of bees left in the colony, so that some may begin purely foraging once most of the hive is recruited.

Our work demonstrates the importance of considering more detailed properties of social insect movement within nests when initiating colony-wide events. The geometry of the nest can distinctly impact the communication and movement strategy that is most efficient for both rapid recruitment and the prevention of false alarms. The modeling framework we have presented here is generalizable to other social animals including social insects like ants~\citep{sumpter03}, but also mammals that initiate group foraging from collective home sites like primates and bats~\citep{janson90,wilkinson98}. Incorporating these additional details into models of collective foraging will better help pin down the learned behavioral mechanisms by which organisms forage efficiently in groups.

\section*{Acknowledgements and Code Availability} SB and ZPK were supported by an NIH (R01MH115557) and NSF (DMS-1853630) grants. SB was also supported by a Dissertation Fellowship from the American Association of University Women. \\

\noindent
Code for producing figures is available at \url{https://github.com/sbidari/hivegeom}

\end{document}